\documentclass[reprint,amsmath,amssymb,aps,superscriptaddress,nofootinbib,floatfix]{revtex4-2}

\usepackage{graphicx}
\usepackage{dcolumn}
\usepackage{bm}
\usepackage{caption}
\usepackage{subcaption}
\usepackage{hyperref}
\usepackage{xcolor}
\usepackage[usestackEOL]{stackengine}

\begin{document}
\strutlongstacks{T}
\newcommand{\St}[1]{\Centerstack{#1}}

\title{Nature of the $N^*$ and $\Delta$ resonances via coupled-channel dynamics}
\author{Yu-Fei Wang}
\email{yuf.wang@fz-juelich.de}
\affiliation{Institute for Advanced Simulation and J\"ulich Center for Hadron Physics, Forschungszentrum J\"ulich, 
52425 J\"ulich, Germany}

\author{Ulf-G.~Mei\ss ner }
\affiliation{Helmholtz-Institut f\"ur Strahlen- und Kernphysik (Theorie) and Bethe Center for Theoretical
Physics,  Universit\"at Bonn, 53115 Bonn, Germany}
\affiliation{Institute for Advanced Simulation and J\"ulich Center for Hadron Physics, Forschungszentrum J\"ulich, 
52425 J\"ulich, Germany}
\affiliation{Tbilisi State University, 0186 Tbilisi, Georgia}

\author{Deborah R{\"o}nchen}
\affiliation{Institute for Advanced Simulation and J\"ulich Center for Hadron Physics, Forschungszentrum J\"ulich, 
52425 J\"ulich, Germany}

\author{Chao-Wei Shen}
\affiliation{Institute for Advanced Simulation and J\"ulich Center for Hadron Physics, Forschungszentrum J\"ulich, 
52425 J\"ulich, Germany}

\begin{abstract}
This work aims at determining the composition of certain $N^*$ and $\Delta$ resonances, i.e. whether they
are compact states formed directly by quarks and gluons, or hadronic molecules generated from the meson-baryon interaction. The information of
the resonance poles is provided by a comprehensive coupled-channel approach, the J\"{u}lich-Bonn model.  $13$
states that are significant in this approach are studied.
Two criteria for each state are adopted in this paper, the comparison thereof roughly
indicates the model uncertainties. It is found that the conclusions for $8$ resonances are relatively certain:
$N(1535) \frac{1}{2}^-$, $N(1440) \frac{1}{2}^+$, $N(1710) \frac{1}{2}^+$, and $N(1520) \frac{3}{2}^-$ tend
to be composite; whereas $N(1650) \frac{1}{2}^-$, $N(1900) \frac{3}{2}^+$, $N(1680) \frac{5}{2}^+$, and
$\Delta(1600) \frac{3}{2}^+$ tend to be compact. 
\end{abstract}
\maketitle
\section{Introduction}
Investigating the inner structures of particles is always an important topic in physics. Especially, the study
of the composition is a fundamental and difficult task, which usually needs firmly established observations
on both the theoretical and the experimental sides, just like the Rutherford scattering experiment which clarified the structure of atoms. In hadron physics, such studies are even more complicated due to the extremely involved dynamics and the fact that most states are instable. 

However, in the 1960ties  Weinberg's work on the deuteron~\cite{weinberg1965} showed that the information of
the composition sometimes could be derived from simple observations. Particularly, if $Z(\in[0,1])$ is the probability
to find an elementary state in the deuteron (the ``elementariness''), then the scattering length $a$ and the
effective range $r$ are
\begin{equation}\label{Weinbergct}
\begin{split}
    &a=-\frac{2(1-Z)}{2-Z}R+\mathcal{O}(L)\ ,\\
    &r=-\frac{Z}{1-Z}R+\mathcal{O}(L)\ ;
\end{split}
\end{equation}
where $R=(2\mu B)^{-1/2}\simeq 4.3$ fm is the binding radius, with $\mu$ the two-nucleon reduced mass and
$B\simeq 2.22$~MeV the deuteron binding energy. Further, $L = 1/M_\pi \simeq 1.4$~fm, with $M_\pi$ the pion mass,
is the typical interaction range of two nucleons. Experiments give $a=-5.41$~fm and $r=+1.75$~fm, which support
a rather small value of $Z$, i.e. the deuteron should be composed by two nucleons. The derivation of this
criterion is straightforward, one just needs to assume a simple formalism including two-nucleon continuous states $|\alpha\rangle_{2N}$ and possible bare states $|n\rangle$ orthogonal to the former, with the coupling constant $g$ between the physical deuteron $|d\rangle$ and the two-nucleon states $|\alpha\rangle_{2N}$. Then the scattering can be solved resulting in the projection probability $Z\equiv 1-\int d\alpha|\langle d|\alpha\rangle_{2N}|^2$. Apart from the original paper~\cite{weinberg1965}, Ref.~\cite{vanKolck:2022lqz} has given a modern explanation of Weinberg's work and shown that it is compatible with the basic conceptions of modern effective field theories (EFTs). For an introduction to EFTs, see~\cite{Meissner:2022cbi}. In addition, Ref.~\cite{Oller:2017alp}  interprets Weinberg's criterion via the number operators of the initial or final state particles, in order to make the connections to the EFTs without bare states. 
Nevertheless, the success of this criterion on the deuteron is somewhat accidental, in the sense that it can hardly
be applied to other states without model assumptions or approximations: the particle must be an $S$-wave stable
bound state near a two-body threshold. 

Several decades after Weinberg's work, many new hadronic states are found. Some of them cannot be interpreted by
the naive quark model, indicating possibly the mechanism of the molecular states formed by the residual hadron-hadron interactions, e.g. hadron exchanges. For example, in  pion-nucleon reactions, there are discussions on the non-trivial structure of the $N^*(1535)$ and $N^*(1440)$ (the Roper resonance), see e.g. Refs.~\cite{Kaiser:1995cy,Glozman:1995fu,Glozman:1996wq,Inoue:2001ip,Nieves:2001wt,Krehl:1999km,Doring:2009uc,Bruns:2010sv,Segovia:2015hra,Wang:2017cfp}. As for
the heavy hadrons, after the discovery of the $X(3872)$~\cite{X3872}, more and more hadron exotic states were found
that deviate significantly from the predictions of the conventional quark model, which can likely be interpreted
by the picture of hadronic molecules~\cite{Guo:2017jvc}~\footnote{In general, one is dealing  with ``dynamically generated'' states, which
are due to the hadron-hadron (or even three-hadron) interactions. Hadronic molecules are a subclass of this type of states close to one or
in between two close-by thresholds.}. A criterion of the composition is urgently needed for
such states. Unfortunately, they are all unstable and Weinberg's criterion cannot be applied directly. 

There are mainly three ways to establish an extended criterion. The first is the ``pole counting rule'' proposed in
Ref.~\cite{Morgan:1992ge}, and applied to various studies of the exotic states, see e.g.
Refs.~\cite{Zhang:2009bv,Dai:2012pb,Meng:2014ota,Gong:2016hlt,Cao:2019wwt,Cao:2020gul,Chen:2021tad}. It abandons
the definition of the unstable state and a ``probability'' as the output, but rather focuses on the pole
structure and the dynamics. A typical quantum mechanical potential of an $S$-wave can only produce one near-threshold
pole, in contrast to the mechanism of Castillejo-Dalitz-Dyson~\cite{Castillejo:1955ed}, which produces two
poles near the threshold. This method is model-independent, but can only be applied to $S$-wave near-threshold states. 

The second method, called ``spectral density function'' approach, was proposed in Ref.~\cite{Baru:2003qq} and further
applied in Refs.~\cite{Kalashnikova:2005ui,Kalashnikova:2009gt,Baru:2010ww,Hanhart:2010wh,Hanhart:2011jz,Gong:2016hlt,Chen:2021tad}. This method also avoids the definition of unphysical states. It has been observed that when lower
channels are switched on, a previously stable state gains a decay width and its elementariness $Z$ disperses
into a finite probability distribution $w(z)$ of the physical energy $z$, called the ``spectral density function''.
Mathematically, $w(z)$ is just the projection of the physical scattering state (with energy $z$) on the bare
elementary state. The elementariness for the resonance can be obtained just by collecting the function $w(z)$
near the pole mass $M_R$ of the resonance: $Z\simeq \int_{M_R-\Delta E}^{M_R+\Delta E}w(z)dz$, with $\Delta E$ a
quantity comparable with the pole width. The mathematical formalism can be constructed in principle for any
partial wave, and the output is quantitative. The choice of $\Delta E$ further generates some
uncertainty, and this method does not work well when the resonance is broad or the overlaps of different resonances are large. 

The third way is defining a quantity similar to the $Z$ in Weinberg's case, the projection of a Gamow
state~\cite{Gamow:1928aa,CIVITARESE200441} on the bare state. The Gamow states describing resonances are
zero-norm~\cite{ROMO1968617} and the corresponding $Z$ is complex, without the interpretation as a ``probability''.
Studies on relevant mathematical properties can be found in, e.g., Refs.~\cite{Xiao:2016dsx,Xiao:2016wbs,Xiao:2016mon}.
Some mathematical transformations or naive measures~\cite{Aceti:2012dd,Hyodo:2013iga,Guo:2015daa,Sekihara:2015gvw,Sekihara:2016xnq,Oller:2017alp,Sekihara:2021eah,Matuschek:2020gqe} are performed to make it a real number between $0$ and $1$. 

In this work, we study the nature of some selected $N^*$ and $\Delta$ states,
based on the resonance parameters extracted in a recent analysis within the J\"{u}lich-Bonn (J\"uBo) dynamical coupled-channel approach~\cite{Ronchen:2022hqk}. See Refs.~\cite{Schutz:1994wp,Schutz:1994ue,Schutz:1998jx,Krehl:1999km,Gasparyan:2003fp,Doring:2009yv,Doring:2010ap,Ronchen:2012eg,Ronchen:2014cna,Ronchen:2015vfa,Ronchen:2018ury,Mai:2021vsw,Mai:2021aui,Shen:2017ayv,Wang:2022oof,Wang:2022osj} for earlier works and other recent developments. In the J\"uBo model, a coupled-channel scattering
equation is solved with angular momentum up to $J=9/2$ and the channel space $\pi N$, $\pi\pi N$ (simulated by three
effective channels $\sigma N$, $\rho N$, and $\pi\Delta$), $\eta N$, $K\Lambda$, and $K\Sigma$. Photoproduction reactions are also taken into account. Ref.~\cite{Wang:2022osj}
further includes the $\pi N\to\omega N$ channel. The free parameters are
determined by the fit to a large collection of pion- and photon-induced data. The analytical
structure is respected and the states
are extracted by means of modern pole-searching procedures in the complex energy plane. Such a model
provides the basic information one needs to study the elementarinesses of resonances. 

Note that many of the $N^*$ and $\Delta$ states are not in $S$-wave, and some of them are not narrow or not
close to any two-particle thresholds. We have to admit that a systematic and accurate study of their nature is
a very difficult task, and it is impossible to completely get rid of the model-dependence. The reason why we
still carry out this study is that the J\"uBo model is data-driven as the parameters are constrained by a
tremendous amount of experimental input. Therefore we believe the findings with respect to the elementarinesses
make sense. Specifically, the pole counting rule cannot be applied in this study, whereas the spectral
density functions approach can be instructive, since these are readily extracted from the $s$-channel bare states
of this model, and require no further assumptions or mathematical transformations. Such functions directly
obtained from this model also measure the
correlation between the input $s$-channel states and output poles, helping us to gain a better understanding on
the interplay of the different components of the model. We also try our best to further estimate the uncertainties by locally constructing another
spectral density function for each state from the pole positions and on-shell residues and comparing it to the
function directly given by the model. Additionally, the Gamow states and the complex $Z$ values are also calculated
as a further check. This is meaningful since
the conception of Gamow states is theoretically very different to the spectral density functions. 

The structure of this paper is as follows. In Sect.~\ref{sec:toy} a solvable toy model is discussed in order
to show the basic concepts and interpretation of the spectral density functions and the
Gamow states. In Sect.~\ref{sec:JB} we briefly summarize the J\"{u}lich-Bonn model and discuss the
spectral density functions and Gamow states in this framework. The numerical results are shown in
Sect.~\ref{sec:result} together with pertinent discussions. Finally Sect.~\ref{sec:con} contains
the conclusions and provides an outlook.

\section{Toy model discussions}\label{sec:toy}
\subsection{Basic descriptions}
To better understand the formalism employed later, we first consider a  toy model. It is non-relativistic and
the reduced mass of the two particles is denoted as $\mu$. Further, the dispersion relation is $\mathcal{E}=k^2/(2\mu)$,
with $k$ the momentum and the threshold is located at the energy $\mathcal{E}=0$. The model contains continuous
free two-body states $|\psi(\mathbf{k})\rangle$ and one isolated bare state $|\psi_0\rangle$ with the energy $E_0$.
For simplicity we assume the bare state to be above the threshold, i.e. $E_0>0$. These free states form an
orthogonal and complete basis: 
\begin{equation}\label{pedabasisdef}
\begin{split}
    &\langle\psi_0|\psi(\mathbf{k})\rangle=\langle\psi(\mathbf{k})|\psi_0\rangle=0\ ,\\
    &\langle\psi_0|\psi_0\rangle=1\ ,\\
    &\langle\psi(\mathbf{k}')|\psi(\mathbf{k})\rangle=(2\pi)^3\delta^{(3)}(\mathbf{k}-\mathbf{k}')\ ,\\
    &|\psi_0\rangle\langle\psi_0|+\int\frac{d^3 \mathbf{k}}{(2\pi)^3}|\psi(\mathbf{k})\rangle\langle\psi(\mathbf{k})|=\hat I\ ,
\end{split}
\end{equation}
where $\hat{I}$ is the identity operator. Any physical state $|\Phi\rangle$ can be expressed as a
linear combination of the free states with the coefficients $c_0$ and $\chi(\mathbf{k})$: 
\begin{equation}\label{pedasolstadef}
    |\Phi\rangle=c_0|\psi_0\rangle+\int\frac{d^3 \mathbf{k}}{(2\pi)^3}\chi(\mathbf{k})|\psi(\mathbf{k})\rangle\ .
\end{equation}

The Hamiltonian consists of a free part $\hat{H}_0$ and an interaction part $\hat{H}_I$, the latter
only contains the $S$-wave coupling between the continuous and the bare states: 
\begin{equation}
\begin{split}
    &\hat{H}_0=E_0|\psi_0\rangle\langle\psi_0|
    +\int\frac{d^3 \mathbf{k}}{(2\pi)^3} \frac{k^2}{2\mu}|\psi(\mathbf{k})\rangle\langle\psi(\mathbf{k})|\ ,\\
    &\hat{H}_I=\int \frac{d^3 \mathbf{k}}{(2\pi)^3}\ gF(k,\Lambda)
    \left[|\psi(\mathbf{k})\rangle\langle \psi_0 |+| \psi_0 \rangle\langle \psi(\mathbf{k}) |\right]\ ; 
\end{split}
\end{equation}
with $g$ the coupling constant and $F(k,\Lambda)$  the regulator
\begin{equation}\label{pedaFexp}
    F(k,\Lambda)=\frac{\Lambda^2}{k^2+\Lambda^2}\ ; 
\end{equation}
and $\Lambda$ is the cut-off parameter. In the following discussions, $\Lambda$ is often assumed to be
significantly larger than any other quantity with energy dimension one, so that we can perform expansions. 
One can also define the free eigenstate with energy $\mathcal{E}$ accordingly: 
\begin{equation}\label{pedaEstadef}
    |\psi(\mathcal{E})\rangle\equiv
    \int\frac{d^3 \mathbf{k}}{(2\pi)^3}\sqrt{\frac{2\pi^2}{\mu k}}
    \delta\left(\frac{k^2}{2\mu}-\mathcal{E}\right)|\psi(\mathbf{k})\rangle\ ,
\end{equation}
which satisfies 
$\hat{H}_0|\psi(\mathcal{E})\rangle=\mathcal{E}|\psi(\mathcal{E})\rangle$. 
In fact this toy model is equivalent or quite similar to many studies based on the spectral density
functions in the literature~\cite{Baru:2003qq,Kalashnikova:2005ui,Kalashnikova:2009gt,Baru:2010ww,Hanhart:2010wh,Hanhart:2011jz,Gong:2016hlt,Chen:2021tad}. 

\subsection{Physical solutions}
The central equation of the toy model is
\begin{equation}\label{pedaphiequ}
    \big(\hat{H}_0+\hat{H}_I\big)|\Phi(\mathcal{E})\rangle=\mathcal{E}|\Phi(\mathcal{E})\rangle\ ,
\end{equation}
which yields 
\begin{equation}\label{pedacchiequ}
\begin{split}
	&E_0c_0(\mathcal{E})+\int \frac{d^3 \mathbf{k}}{(2\pi)^3}\ gF(k,\Lambda)\chi(\mathcal{E},\mathbf{k})
    =\mathcal{E}c_0(\mathcal{E})\ ,\\
	&\frac{k^2}{2\mu}\chi(\mathcal{E},\mathbf{k})+gF(k,\Lambda)c_0(\mathcal{E})=\mathcal{E}\chi(\mathcal{E},\mathbf{k})\ . 
\end{split}
\end{equation}
Note that the coefficients depend on the energy eigenvalue $\mathcal{E}$. 

\subsubsection{Bound states}
A bound state corresponds to $\mathcal{E}=-B<0$, with the binding energy $B$. In this case the coefficient
$\chi$ can be eliminated by the second line of Eq.~\eqref{pedacchiequ}. Then, the coefficient $c_0$ in the
first line of Eq.~\eqref{pedacchiequ} is also eliminated, leaving only one equation for the binding energy: 
\begin{equation}\label{pedaSigdef}
\begin{split}
    &-B-E_0-\Sigma(-B)=0\ ,\\
    &\Sigma(\mathcal{E})\equiv
    \int\frac{d^3\mathbf{k}}{(2\pi)^3}\frac{g^2F^2(k,\Lambda)}{\mathcal{E}-k^2/(2\mu)}\ ,
\end{split}
\end{equation}
where $\Sigma(\mathcal{E})$ is the self-energy of the bare state. The energy of the bound state is
a pole of the propagator
\begin{equation}\label{pedaprop}
    D(\mathcal{E})\equiv \frac{1}{\mathcal{E}-E_0-\Sigma(\mathcal{E})}\ .
\end{equation}
The bound state wave function is written as: 
\begin{equation}\label{pedabs}
    |\Phi(-B)\rangle=c_0\left[|\psi_0\rangle
    -\int\frac{d^3 \mathbf{k}}{(2\pi)^3}\frac{gF(k,\Lambda)}{\frac{k^2}{2\mu}+B}|\psi(\mathbf{k})\rangle\right]\ .
\end{equation}
The coefficient $c_0$ can further be determined by the normalization condition: 
\begin{equation}\label{pedaBnorm}
    \langle \Phi(-B)|\Phi(-B)\rangle=1\ ,
\end{equation}
which yields the elementariness
\begin{equation}\label{pedaZBsol}
    Z\equiv|c_0|^2=\frac{1}{1-\Sigma'(-B)}\ ,
\end{equation}
where the prime denotes differentiation with respect to energy.
Moreover, if one expands to leading order in the inverse of the large cut-off $\Lambda$, the elementariness can be
written as
\begin{equation}\label{pedaZBexp}
    Z=\left(1+\frac{g^2\mu^2}{2\pi\sqrt{2\mu B}}\right)^{-1}+\mathcal{O}(\Lambda^{-1})\ ;
\end{equation}
which exhibits the typical feature $g^2\propto (Z^{-1}-1)$ as discussed in Ref.~\cite{Guo:2017jvc}.
The compositeness is further defined as
\begin{equation}
    X\equiv \int\frac{d^3 \mathbf{k}}{(2\pi)^3}|\chi(-B,\mathbf{k})|^2\ ,
\end{equation}
so that according to the normalization condition Eq.~\eqref{pedaBnorm}, $Z+X=1$. 

\subsubsection{Scattering states}
For the scattering states, any $\mathcal{E}>0$ can be an eigenvalue. In this case one should give the
energy an infinitesimal imaginary part: $\mathcal{E}\to \mathcal{E}+i0^+$. It is convenient to eliminate
the coefficient $c_0$ by the first line of 
Eq.~\eqref{pedacchiequ}, and then the second line becomes
\begin{equation}\label{pedawavef}
\begin{split}
    &\int\frac{d^3\mathbf{p}}{(2\pi)^3}
    \Big[\frac{p^2}{2\mu}(2\pi)^3\delta^{(3)}(\mathbf{p}-\mathbf{k})\\
    &+\frac{g^2F(k,\Lambda)F(p,\Lambda)}{\mathcal{E}-E_0+i0^+}\Big]\chi(\mathcal{E},\mathbf{p})
    =\mathcal{E}\chi(\mathcal{E},\mathbf{k})\ .
\end{split}
\end{equation}
Here, $\chi$ can be understood as the wave function in momentum space. The corresponding
potential operator and the Hamiltonians are 
\begin{equation}\label{pedaV}
\begin{split}
    &\hat{V}=\hat{H}_I\frac{|\psi_0\rangle \langle\psi_0|}{\mathcal{E}-E_0+i0^+}\hat{H}_I\ ,\\
    &\hat{h}_0=\int\frac{d^3 \mathbf{k}}{(2\pi)^3} \frac{k^2}{2\mu}|\psi(\mathbf{k})\rangle\langle\psi(\mathbf{k})|\ ,\\
    &\hat{h}\equiv \hat{h}_0+\hat{V}\ .
\end{split}
\end{equation}
The scattering amplitude can be worked out with the help of the Lippmann-Schwinger equation
\begin{equation}\label{pedaTequ}
    \hat{T}=\hat{V}+\hat{V}\hat{G}\hat{T}\ ,\quad \hat{G}\equiv (\mathcal{E}-\hat{h}_0+i0^+)^{-1}\ .
\end{equation}
The scattering amplitude is found as
\begin{equation}\label{pedaamp}
    \langle\psi(\mathbf{k})|\hat{T}|\psi(\mathbf{p})\rangle\equiv T(k,p,\mathcal{E})
    =\frac{g^2F(k,\Lambda)F(p,\Lambda)}{\mathcal{E}-E_0-\Sigma(\mathcal{E}+i0^+)}\ ,
\end{equation}
which corresponds to an $s$-channel Feynman diagram with the dressed propagator. The physical (on-shell)
scattering amplitude should be $T(q_\varepsilon,q_\varepsilon,\mathcal{E})$, shortened as $T(\mathcal{E})$,
with $q_\varepsilon=\sqrt{2\mu \mathcal{E}}$ the on-shell momentum.

It is meaningful to relate $\chi$ to the amplitude. The definition of the $\hat{\chi}$ operator is: 
\begin{equation}
    \chi(\mathcal{E},\mathbf{k})\equiv \langle\psi(\mathbf{k})|\hat{\chi}|\psi(\mathcal{E})\rangle\ ,
\end{equation}
then Eq.~\eqref{pedawavef} is translated as
\begin{equation}\label{pedachiequ}
    \hat{V}\hat{\chi}|\psi(\mathcal{E})\rangle=\hat{G}^{-1}\hat{\chi}|\psi(\mathcal{E})\rangle\ .
\end{equation}
One may also rewrite the scattering equation Eq.~\eqref{pedaTequ}, considering $\hat{G}^{-1}|\psi(\mathcal{E})\rangle=0$: 
\[
\hat{G}^{-1}(\hat{I}+\hat{G}\hat{T})|\psi(\mathcal{E})\rangle
=\hat{V}(\hat{I}+\hat{G}\hat{T})|\psi(\mathcal{E})\rangle\ ;
\]
then
\begin{equation}\label{pedachisolop}
    \hat{\chi}|\psi(\mathcal{E})\rangle
    =\mathcal{N}(\mathcal{E})(\hat{I}+\hat{G}\hat{T})|\psi(\mathcal{E})\rangle\ ,
\end{equation}
with $\mathcal{N}$ a normalization factor that might depend on $\mathcal{E}$. Finally the solution of $\chi$ is
\begin{equation}\label{pedachisol}
\begin{split}
    \chi(\mathcal{E},\mathbf{k})&=\mathcal{N}(\mathcal{E})
    \Bigg[\sqrt{\frac{2\pi^2}{\mu k}}\delta\Big(\frac{k^2}{2\mu}-\mathcal{E}\Big)\\
    &+\sqrt{\frac{\mu q_\varepsilon}{2\pi^2}}\frac{T(k,q_\varepsilon,\mathcal{E})}{\mathcal{E}-k^2/(2\mu)+i0^+}\Bigg]\ .
\end{split}
\end{equation}
According to Eq.~\eqref{pedacchiequ},
\begin{equation}
    c_0(\mathcal{E})=\mathcal{N}(\mathcal{E})\sqrt{\frac{\mu q_\varepsilon}{2\pi^2}}
    \frac{T(k,q_\varepsilon,\mathcal{E})}{gF(k,\Lambda)}\ .
\end{equation}
The normalization factor is determined as $\mathcal{N}(\mathcal{E})=1$ by the normalization condition
of the scattering states
\begin{equation}\label{pedascnorm}
    \langle \Phi(\mathcal{E}')|\Phi(\mathcal{E})\rangle=\delta(\mathcal{E}-\mathcal{E}')\ .
\end{equation}
The spectral density function is defined as the probability density of finding the bare state
in the scattering states: 
\begin{equation}\label{pedasdfamp}
\begin{split}
    w(\mathcal{E})&=|c_0(\mathcal{E})|^2\\
    &=\rho
    \Bigg|\frac{T(k,q_\varepsilon,\mathcal{E})}{gF(k,\Lambda)}\Bigg|^2
    =\rho
    \Bigg|\frac{g F(q_{\varepsilon},\Lambda)}{\mathcal{E}-E_0-\Sigma(\mathcal{E})}\Bigg|^2\ , 
\end{split}
\end{equation}
with the factor $\rho=\mu q_{\varepsilon}/(2\pi^2)$. 

Note that the effective range expansion can be performed on Eq.~\eqref{pedaamp}. Combining with Eq.~\eqref{pedaZBexp},
Weinberg's criterion Eq.~\eqref{Weinbergct} can be reproduced by this toy model. 

\subsection{Poles and criteria}
\subsubsection{Pole trajectories}
Eqs.~\eqref{pedaSigdef} and \eqref{pedaamp} indicate that the bound states are just the poles of the physical
scattering amplitude in the energy region $\mathcal{E}<0$. Besides, the dispersion relation $q_{\varepsilon}=
\sqrt{2\mu\mathcal{E}}$ divides the complex energy plane into two Riemann sheets by the branch cut
$\mathcal{E}\in[0,+\infty)$, with the physical sheet featured as $\text{Im}\,q_{\varepsilon}>0$. The unphysical
states appear as poles on the second sheet with $\text{Im}\,q_{\varepsilon}<0$. Among those poles, virtual states
lie on the real energy axis below the threshold, while the resonances have non-zero imaginary parts. 

\begin{figure*}[htbp]
	\centering
	\begin{subfigure}[htbp]{0.48\textwidth}
		\centering
		\includegraphics[width=\textwidth]{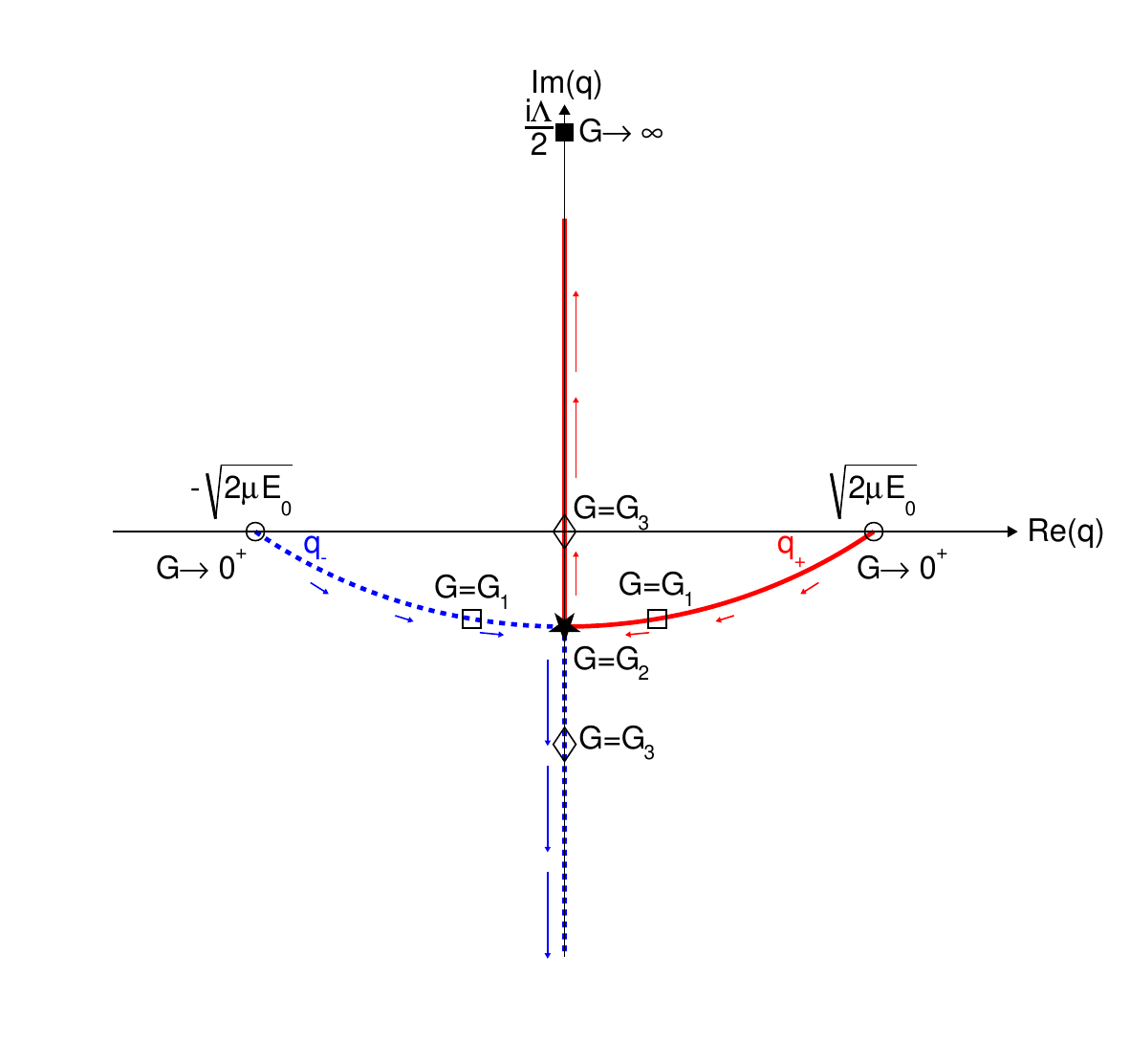}
		\caption{the momentum plane}
		\label{fig:pedapole:p}
	\end{subfigure}
	\hfill
	\begin{subfigure}[htbp]{0.48\textwidth}
		\centering
		\includegraphics[width=\textwidth]{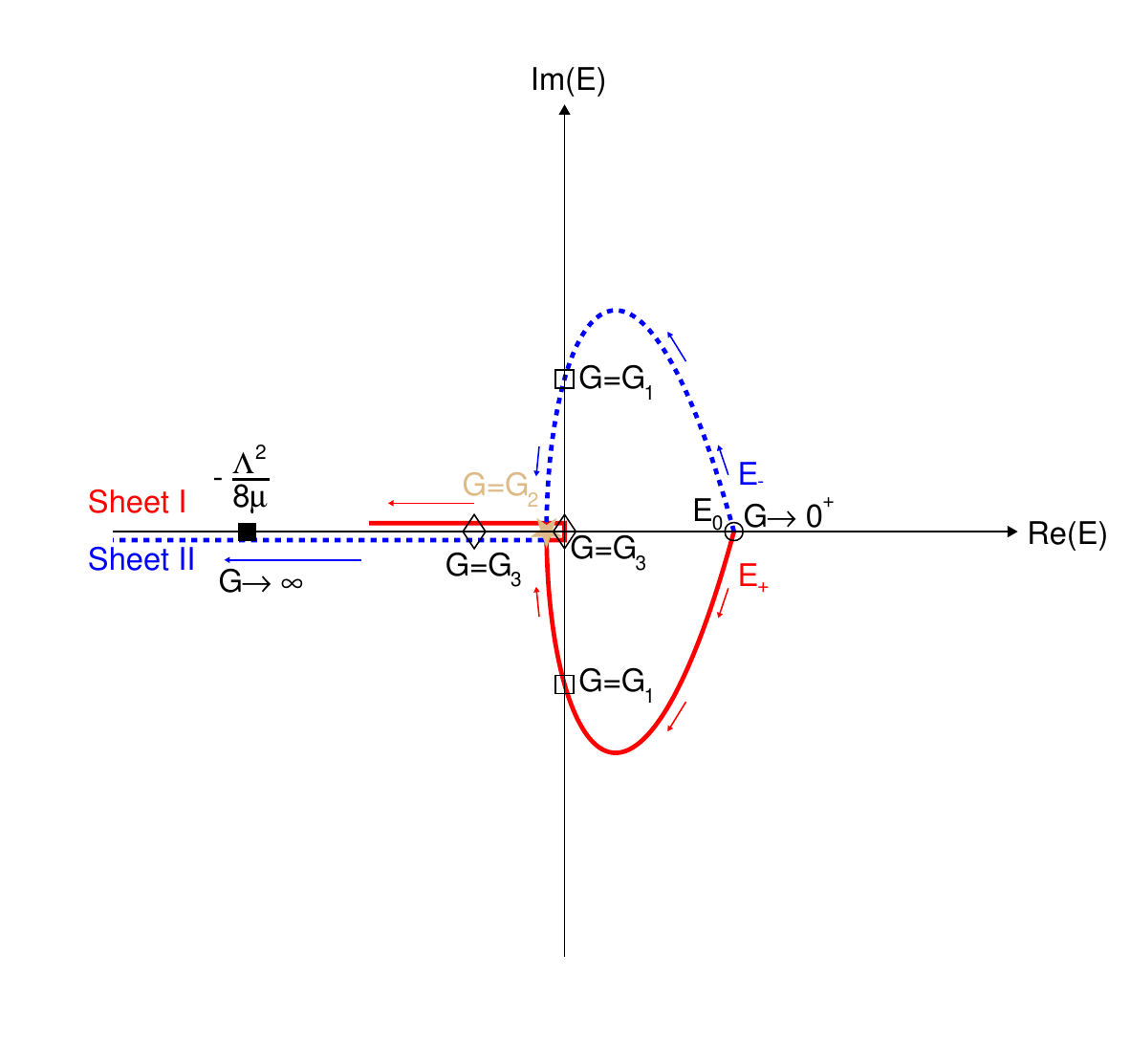}
		\caption{the energy plane}
		\label{fig:pedapole:e}
	\end{subfigure}
	\caption{The schematic plot of the pole trajectories in the toy model. The red solid (blue dashed) line denotes the trajectory of the first (second) pole. The critical points are as follows.
          $G_1$: a resonance pole with the zero width. $G_2$: a resonance pole and the conjugate pole collide,
          forming a second-order virtual state pole. $G_3$: a bound state is produced. }
	\label{fig:pedapole}
\end{figure*}

We can expand the denominator of the amplitude in Eq.~\eqref{pedaamp} to leading order in the inverse of the large cut-off $\Lambda$, which gives a result similar to
the Flatt{\'e} parametrization: 
\begin{equation}\label{pedaTexp}
    T(\mathcal{E})
    =\frac{g^2}{\mathcal{E}-E_0+\frac{g^2\mu\Lambda}{4\pi}+i\frac{g^2\mu}{2\pi}q_\varepsilon}
    +\mathcal{O}(\Lambda^{-1})\ .
\end{equation}
Defining 
\begin{equation}
    G\equiv g^2\mu^2/(2\pi)\ ,
\end{equation}
the trajectories of the poles\footnote{Note that $G\to\infty$ is not
consistent with the expansion in $1/\Lambda$. However, we tentatively keep a finite value of $\Lambda$ to
finish the discussion, which can be regarded as another model based on the Flatt{\'e} parametrization. }
can be plotted, see Fig.~\ref{fig:pedapole}. 
For similar discussions see e.g. Refs.~\cite{Guo:2017jvc,Hanhart:2014ssa}. 
Under the condition $E_0>0$, the bound states exists only if the following condition holds:  
\begin{equation}
    G>G_3\equiv \frac{2\mu E_0}{\Lambda}\ .
\end{equation}

\subsubsection{Spectral density function and  poles}
The physical solutions of Eq.~\eqref{pedaphiequ} also form a complete and orthogonal basis: 
\begin{equation}\label{pedasolcom}
    |\Phi(-B)\rangle\langle\Phi(-B)|+\int_0^{+\infty}d\mathcal{E}|\Phi(\mathcal{E})\rangle\langle\Phi(\mathcal{E})|
    =\hat I\ ,
\end{equation}
which, by applying to the bare state $\langle\psi_0|\cdots|\psi_0\rangle$, leads to
\begin{equation}\label{pedasr}
    Z+\int_0^\infty d\mathcal{E} w(\mathcal{E})=1\ ,
\end{equation}
where $Z$ is the elementariness of the bound state. This is the spectral density function sum rule, which
indicates the free bare state will surely be found if one explores every physical solution. It is
also worth mentioning that the spectral density function is proportional to the imaginary part of the
propagator in Eq.~\eqref{pedaprop}: 
\begin{equation}\label{pedawim}
    w(\mathcal{E})=-\frac{1}{\pi}\text{Im}D(\mathcal{E})\ .
\end{equation}
Therefore the sum rule Eq.~\eqref{pedasr} can also be understood as the K{\"a}ll{\'e}n–Lehmann spectral
representation of an isolated state. 

To make closer contact to the physics, we investigate the spectral density function near the pole enery of
a narrow resonance. Taking a small value of $g$ in Eq.~\eqref{pedaTexp}, and suppressing the
quantities with $g^4$ and $\Lambda^{-1}$, the pair of resonance poles is located at: 
\begin{equation}
\begin{split}
    &\mathcal{E}_\pm=\left(E_0-\frac{G\Lambda}{2\mu}\right)\mp \frac{i G\sqrt{2\mu E_0}}{\mu}\\
    &\equiv \mathcal{E}_R\mp\frac{i}{2}\Gamma_R\ ;
\end{split}
\end{equation}
where $\mathcal{E}_R$ and $\Gamma_R$ denote the pole energy and pole width,
respectively. According to Eq.~\eqref{pedasdfamp}, the spectral density function near the pole energy is
\begin{equation}\label{pedaBWw}
    w(\mathcal{E}\simeq \mathcal{E}_R)=\frac{1}{\pi}\frac{\Gamma_R/2}{(\mathcal{E}-\mathcal{E}_R)^2+\Gamma_R^2/4}
    \ .
\end{equation}
In other words, 
\begin{equation}\label{pedaBWlim}
    \lim_{g\to 0}w(\mathcal{E}\simeq \mathcal{E}_R)=\delta(\mathcal{E}-\mathcal{E}_R)\ .
\end{equation}
On the one hand, as $g\to 0$ the bare state decouples from the continuous states, and the Hamiltonian
becomes free, so then the state at $\mathcal{E}_R=E_0$ has elementariness $Z=1$. On the other hand,
the spectral density function is concentrated totally on a single energy point. Since a finite $g$ just makes
the $\delta$ function disperse like Eq.~\eqref{pedaBWw}, we can still assume the spectral density function
near $\mathcal{E}=\mathcal{E}_R$ being a measure of the elementariness of the resonance, which is the central
idea for applying this method. Nevertheless, one also finds that the broader the resonance is, the more ambiguous
this method will be. In fact, when the resonance is too broad even Eq.~\eqref{pedaBWw} becomes problematic. 

\subsubsection{Gamow states}
\begin{figure}[t!]
	\centering
	\includegraphics[width=0.3\textwidth]{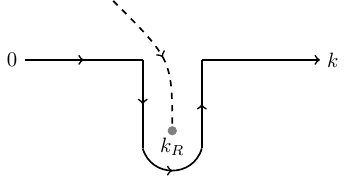}
	\caption{Reaching the second sheet by contour deformation. The solid line stands for the integral contour, while the dashed line represents the move of the singularity $k=k_R$ when the energy moves across the cut. See the text for explanations. }
	\label{fig:pedadeform}
\end{figure}

Resonances correspond to the poles on the second Riemann sheet
\begin{equation}\label{pedaERequ}
    E_R-E_0-\Sigma^{\text{II}}(E_R)=0\ ,
\end{equation}
where the self-energy on the second sheet is evaluated by the deformed integral contour $C_+$:
\begin{equation}
\begin{split}
    \Sigma^{\text{II}}(E_R)
    &=\int_{C^+}\frac{k^2dk}{2\pi^2}\frac{g^2F^2(k,\Lambda)}{E_R-k^2/(2\mu)}\\
    &\equiv \int\frac{k^2dk}{2\pi^2}\frac{g^2F^2(k,\Lambda)}{\Big(E_R-\frac{k^2}{2\mu}\Big)_+}\ .
\end{split}
\end{equation} 
To reach the second sheet, the energy $E_R$ moves across the cut $\mathcal{E}\in [0,\infty)$, and
meanwhile the pole of the integrand $k_R=\sqrt{2\mu E_R}$ hits the original contour $k\in [0,\infty)$ and
deforms it. The deformed contour is topologically the original contour plus a residue term at $k_R=\sqrt{2\mu E_R}$,
see Fig.~\ref{fig:pedadeform}. 
Then the Gamow states can be defined accordingly (see Eq.~\eqref{pedabs}): 
\begin{equation}\label{pedaresst}
\begin{split}
    &|\Phi(E_R))\equiv c_0\Bigg[|\psi_0\rangle
    +\int\frac{d^3 \mathbf{k}}{(2\pi)^3}\frac{gF(k,\Lambda)}{\Big(E_R-\frac{k^2}{2\mu}\Big)_+}|\psi(\mathbf{k})\rangle\Bigg]\ ,\\
    &|\Phi(E_R^*))\equiv c_0^*\Bigg[|\psi_0\rangle
    +\int\frac{d^3 \mathbf{k}}{(2\pi)^3}\frac{gF(k,\Lambda)}{\Big(E_R^*-\frac{k^2}{2\mu}\Big)_-}|\psi(\mathbf{k})\rangle\Bigg]\ ,
\end{split}
\end{equation}
where $E_R^*$ is the conjugate pole and the ``$-$'' refers to the corresponding deformed contour. Note that
the contour deformation in Eq.~\eqref{pedaresst} does not introduce new $|\psi(\mathbf{k})\rangle$'s with
complex $\mathbf{k}$. Instead it only changes the rule of the inner product: whenever another state
$\langle\phi|$ does the inner product with the Gamow state, namely $\langle\phi|\Phi(E_R))$, the integral
contour should be deformed. Eq.~\eqref{pedaERequ} ensures the Gamow state is an eigenstate of the Hamiltonian,
but since the Hamiltonian is Hermitian, the Gamow state cannot be normalized.
The normalization should be defined via the conjugate part: 
\begin{equation}\label{pedaresnorm}
    (\Phi(E_R^*)|\Phi(E_R))=1\ .
\end{equation}
From this condition one can only define the $c_0^2$, instead of $|c_0|^2$, as the elementariness: 
\begin{equation}\label{pedaZgeneral}
    Z_R\equiv c_0^2=\left[1-\frac{d}{d\mathcal{E}}\Sigma^{\text{II}}(\mathcal{E}=E_R)\right]^{-1}\ ,
\end{equation}
which is a complex quantity without the interpretation as a probability. Then the compositeness
can also be defined as $X_R\equiv 1-Z_R$. 

The compositeness of the Gamow state can also be related to the scattering amplitude. On the one hand,
equivalently to Eq.~\eqref{pedaTequ}, the scattering  operator $\hat{T}$ can be rewritten as
\begin{equation}\label{pedaTequ2}
    \hat{T}=\hat{V}+\hat{V}\hat{\mathcal{G}}\hat{V}\ ,\quad \hat{\mathcal{G}}\equiv (\mathcal{E}-\hat{h}+i0^+)^{-1}\ ,
\end{equation}
where $\hat{\mathcal{G}}$ is the full Green's operator of the Hamiltonian $\hat{h}$, see Eq.~\eqref{pedaV}. On
the other hand, when there is a resonance, the completeness condition Eq.~\eqref{pedasolcom} can be modified
\begin{equation}\label{pedasolcommod}
    |\Phi(E_R))(\Phi(E_R^*)|+\int_0^{+\infty}d\mathcal{E}|\tilde{\Phi}(\mathcal{E})\rangle\langle\Phi(\mathcal{E})|
    =\hat I\ ,
\end{equation}
where $|\tilde{\Phi}(\mathcal{E})\rangle$ is the modified scattering state, for details see Ref.~\cite{PETROSKY1991175}.
Since $\hat{\mathcal{G}}|\Phi(E_R))=(\mathcal{E}-E_R)^{-1}|\Phi(E_R))$, the Laurent expansion of the $T$
amplitude on the second sheet can be obtained from the modified completeness condition: 
\begin{equation}
    T^{\text{II}}(k,p,\mathcal{E})
    =\frac{\mathfrak{r}(k)\mathfrak{r}(p)}{\mathcal{E}-E_R}+\cdots\ ;
\end{equation}
with the off-shell residue
\begin{equation}
    \mathfrak{r}(k)\equiv \langle\psi(\mathbf{k})|\hat{V}|\Phi(E_R))\ .
\end{equation}
Moreover, 
$\langle\psi(\mathbf{k})|\hat{V}|\Phi(E_R))=\langle\psi(\mathbf{k})|(\hat{h}-\hat{h}_0)|\Phi(E_R))=[E_R-k^2/(2\mu)]_+
\langle\psi(\mathbf{k})|\Phi(E_R))$, where ``$+$'' denotes the deformed contour, one obtains the
compositeness from the off-shell residue 
\begin{equation}\label{pedaXgeneral}
\begin{split}
    X(E_R)
    &=\int \frac{d^3\mathbf{k}}{(2\pi)^3} \langle\psi(\mathbf{k})|\Phi(E_R))^2\\
    &=\int \frac{d^3\mathbf{k}}{(2\pi)^3}\frac{\mathfrak{r}^2(k)}{\big[E_R-k^2/(2\mu)\big]_+^2}\ .
\end{split}
\end{equation}
According to Eq.~\eqref{pedaamp}, the off-shell residue in this toy model is
$\mathfrak{r}(k)=gF(k,\Lambda)[1-\Sigma^{\text{II}\prime}]^{-1/2}$, so Eq.~\eqref{pedaXgeneral} gives exactly
the same result as Eq.~\eqref{pedaZgeneral}.

\subsection{Coupled-channel extension}
Here, we slightly extend the toy model to a coupled-channel situation. We keep the channel discussed above
as channel ``$1$'', with all the notations remaining the same. In addition, there is a lower channel with
index ``$0$''. The threshold of the lower channel is $-\Delta E<0$. The free continuous states of
channel $0$ are labelled as $|\psi^0(\mathbf{k})\rangle$, which are orthogonal to the $|\psi(\mathbf{k})\rangle$.
For a given energy $\mathcal{E}$, the on-shell momentum of the channel $1$ remains the same, i.e.
$q_\varepsilon=\sqrt{2\mu \mathcal{E}}$, while for channel $0$ it is $q_{0\varepsilon}=\sqrt{2\mu_0
(\mathcal{E}+\Delta E)}$. Equivalently to Eq.~\eqref{pedaEstadef}, we define another energy eigenstate
for the lower channel $0$: 
\begin{equation}\label{pedaEstadef2ch}
\begin{split}
    &|\psi^0(\mathcal{E})\rangle\\
    &=\int\frac{d^3 \mathbf{k}_0}{(2\pi)^3}\sqrt{\frac{2\pi^2}{\mu_0 k_0}}
    \delta\left(\frac{k_0^2}{2\mu_0}-\mathcal{E}-\Delta E\right)|\psi^0(\mathbf{k}_0)\rangle\ .
\end{split}
\end{equation}
Besides, there is still only one bare state $|\psi_0\rangle$. 

The Hamiltonian now is\footnote{The cut-off $\Lambda$ is chosen to be the same for the two channels. } 
\begin{equation}\label{pedaH2ch}
\begin{split}
    \hat{H}_0&=E_0|\psi_0\rangle\langle\psi_0|
    +\int\frac{d^3 \mathbf{k}}{(2\pi)^3} \frac{k^2}{2\mu}|\psi(\mathbf{k})\rangle\langle\psi(\mathbf{k})|\\
    &+\int\frac{d^3 \mathbf{k}_0}{(2\pi)^3} \Big(\frac{k_0^2}{2\mu_0}-\Delta E\Big)
    |\psi^0(\mathbf{k})\rangle\langle\psi^0(\mathbf{k})|\ ,\\
    \hat{H}_I&=\int \frac{d^3 \mathbf{k}}{(2\pi)^3}\ gF(k,\Lambda)
    |\psi(\mathbf{k})\rangle\langle \psi_0 |\\
    &+\int \frac{d^3 \mathbf{k}_0}{(2\pi)^3}\ g_0F(k_0,\Lambda)
    |\psi^0(\mathbf{k}_0)\rangle\langle \psi_0 |+\text{h.c.}\ ; 
\end{split}
\end{equation}
where ``h.c.'' refers to the Hermitian conjugation. The physical scattering state is
\begin{equation}\label{pedasolsta2ch}
\begin{split}
    |\Phi(\mathcal{E})\rangle&
    =c_0(\mathcal{E})|\psi_0\rangle
    +\int\frac{d^3 \mathbf{k}}{(2\pi)^3}\chi(\mathcal{E},\mathbf{k})|\psi(\mathbf{k})\rangle\\
    &+\int\frac{d^3 \mathbf{k}_0}{(2\pi)^3} \chi_0(\mathcal{E},\mathbf{k}_0)|\psi^0(\mathbf{k}_0)\rangle\ .
\end{split}
\end{equation}
The eigenstates can still be solved from Eq.~\eqref{pedaphiequ}. 
In this case, the amplitude is 
\begin{equation}\label{pedaamp2ch}
    T_{ij}(k,p,\mathcal{E})=\frac{g_ig_jF(k,\Lambda)F(p,\Lambda)}{\mathcal{E}-E_0-\widetilde{\Sigma}(\mathcal{E}+i0^+)}\ , 
\end{equation}
with $i,j=0,1$ as the channel indices, and $g_1=g$. Meanwhile the propagator is  
\begin{equation}
    D(\mathcal{E})=\frac{1}{\mathcal{E}-E_0-\widetilde{\Sigma}(\mathcal{E}+i0^+)}\ , 
\end{equation}
with the two-channel self-energy 
\begin{equation}\label{pedaSigdef2ch}
\begin{split}
    \widetilde{\Sigma}(\mathcal{E})
    &=\int\frac{d^3\mathbf{k}}{(2\pi)^3}
    \Big[\frac{g_0^2F^2(k,\Lambda)}{\mathcal{E}-\frac{k^2}{2\mu_0}+\Delta E+i0^+}\\
    &+\frac{g^2F^2(k,\Lambda)}{\mathcal{E}-\frac{k^2}{2\mu}+i0^+}\Big]\ .
\end{split}
\end{equation}

Due to the same reason as in the single channel case, for both $i=0,1$ ($|\psi^1(\mathcal{E})\rangle
=|\psi(\mathcal{E})\rangle$)
\begin{equation}\label{pedachiT2ch}
    \hat{\chi}|\psi^i(\mathcal{E})\rangle
    =\mathcal{N}_i(\mathcal{E})(\hat{I}+\hat{G}\hat{T})|\psi^i(\mathcal{E})\rangle\ .
\end{equation}
The correspondence of the function $\chi$ to the $\hat\chi$ operator is not unique in the coupled-channel system,
and the normalization factor $\mathcal{N}$ is no longer trivially one. Fortunately the spectral density function
is easily obtained from Eq.~\eqref{pedawim}: 
\begin{equation}\label{pedasdf2ch}
    w(\mathcal{E})=
    \frac{g_0^2\mu_0q_{0\varepsilon}F^2(q_{0\varepsilon},\Lambda)
    +g^2\mu q_{\varepsilon}F^2(q_{\varepsilon},\Lambda)\Theta(\mathcal{E})}
    {2\pi^2|\mathcal{E}-E_0-\widetilde{\Sigma}(\mathcal{E})|^2}\ ;
\end{equation}
where $\Theta(\mathcal{E})$ is the Heaviside step function. It is easy to verify that when $g_0=0$,
Eq.~\eqref{pedasdfamp} is recovered. 

At last we investigate a ``quasi-bound state'' ($qb$) in the two-channel case. We assume when channel~$0$ is
switched off, the channel~$1$ forms a bound state at $\mathcal{E}=-B$, with the binding energy $B\ll \Delta E$.
As the coupling $g_0$ grows from zero but is still small, the bound state becomes a narrow resonance. Such a
pole lies on the Riemann sheet defined by $\text{Im}\,q_{0\varepsilon}<0,\text{Im}\,q_{\varepsilon}>0$, i.e.
the physical sheet for channel~$1$ but unphysical sheet for channel~$0$. 
Expanding Eq.~\eqref{pedaSigdef2ch} and suppressing the $\mathcal{O}(g_0^4)$ and $\mathcal{O}(\Lambda^{-1})$ terms,
the location of the pole is (the conjugate one is omitted)
\begin{equation}\label{pedaqbpole}
\begin{split}
    &\mathcal{E}_+=E_{qb}-\frac{i}{2}\Gamma_{qb}\ ,\\
    &E_{qb}=-B-\frac{g_0^2\mu_0\Lambda}{4\pi}Z_B\ ,\\
    &\Gamma_{qb}=\frac{g_0^2\mu_0}{\pi}\sqrt{2\mu_0(\Delta E-B)}Z_B\ ,
\end{split}
\end{equation}
where $Z_B=(1-\Sigma'(-B))^{-1}$ is the elementariness of the bound state when the channel~$0$ is switched off.
Moreover, the spectral density function near the pole energy is 
\begin{equation}\label{pedasdfqb}
    w(\mathcal{E}\simeq E_{qb})=\frac{Z_B}{\pi}
    \frac{\Gamma_{qb}/2}
    {(\mathcal{E}-E_{qb})^2+(\Gamma_{qb}/2)^2}\ .
\end{equation}
Hence in the narrow resonance limit,
\begin{equation}\label{pedaBWlim2ch}
    \lim_{g_0\to 0}w(\mathcal{E}\simeq E_{qb})=Z_B\delta(\mathcal{E}+B)\ .
\end{equation}
This indicates again that the spectral density function near the pole energy carries the information of the
elementariness, which is fully compatible with the elementariness of the bound state (Weinberg's criterion).
When $g_0$ is finite, the delta function in Eq.~\eqref{pedaBWlim2ch} disperses and becomes a finite distribution. 

The elementariness of a quasi-bound state with finite width can be evaluated as $\int_{E_{qb}-\Delta E}^{E_{qb}+\Delta E}
w(\mathcal{E})d\mathcal{E}$. Actually the choice of the integral interval ($\Delta E$) is ambiguous.
Note that the $Z_B$ in Eq.~\eqref{pedasdfqb} is the crucial quantity we are interested in. So for the pole at $\mathcal{E}
=E_R-\frac{i}{2}\Gamma_R$, we define the ``Breit-Wigner'' spectral density function only according to
the pole position: 
\begin{equation}\label{pedabww}
    BW(\mathcal{E})\equiv\frac{1}{\pi}
    \frac{\Gamma_R/2}
    {(\mathcal{E}-E_R)^2+(\Gamma_R/2)^2}\ . 
\end{equation}
Then the elementariness is modified as 
\begin{equation}\label{pedaZReval}
    Z \simeq \frac{\int_{E_R-\Delta E}^{E_R+\Delta E}w(\mathcal{E})d\mathcal{E}}
    {\int_{E_R-\Delta E}^{E_R+\Delta E}BW(\mathcal{E})d\mathcal{E}}\ .
\end{equation}
Since in the narrow resonance limit of Eq.~\eqref{pedasdfqb}, Eq.~\eqref{pedaZReval} gives exactly the quantity
$Z_B$ no matter what value $\Delta E$ is, it is expected that the dependence of the result on $\Delta E$ is
weakened. However, again, when the resonance is too broad, Eq.~\eqref{pedasdfqb} does not hold and this
evaluation is also ambiguous. 

The Gamow states for coupled channles can also be discussed in a totally similar manner to the single channel case.
We skip the details here. 

\section{Application to the J\"{u}lich-Bonn model}\label{sec:JB}
\subsection{Short description of the  model}
\begin{figure*}[t!]
	\centering
	\includegraphics[width=0.7\textwidth]{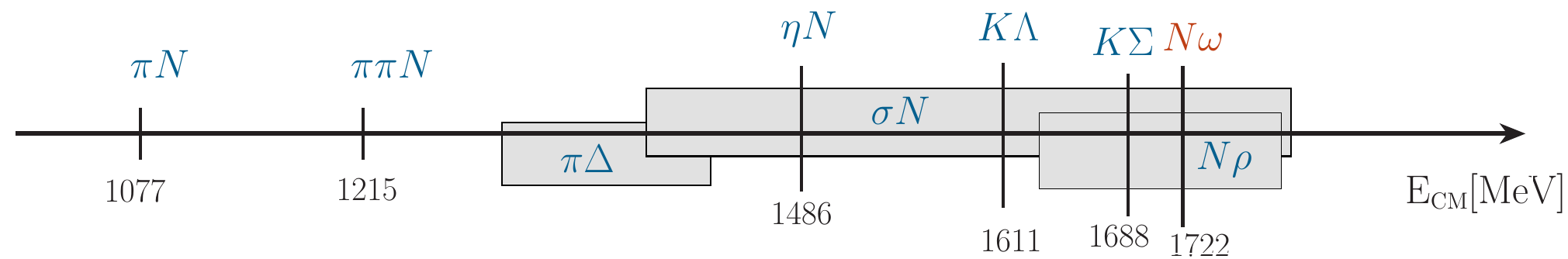}
	\caption{Thresholds of the scattering channels currently considered in the J\"{u}lich-Bonn model as
          function of the center-of-mass energy. The $\omega N$ channel is not considered
          in this work, see the discussions in the beginning of Sect.~\ref{sec:result}. }
	\label{fig:channelthrs}
\end{figure*}

The J\"{u}lich-Bonn model is a comprehensive coupled-channel model, which currently contains the hadronic channels 
$\pi N$, $\pi\pi N$, $\eta N$, $K \Lambda$, $K \Sigma$ and $\omega N$. The $\pi\pi N$ system is
simulated by three effective channels, namely $\pi \Delta$, $\sigma N$ and $\rho N$. The thresholds are
shown in Fig.~\ref{fig:channelthrs}. 

In this model the reactions are studied through the following scattering equation: 
\begin{equation}\label{scequ}
\begin{split}
  &T_{\mu\nu}(p'',p',z)=V_{\mu\nu}(p'',p',z)\\
  &+\sum_{\kappa}\int_0^\infty p^2 dp V_{\mu\kappa}(p'',p,z)G_{\kappa}(p,z)T_{\kappa\nu}(p,p',z)\ ,
\end{split}
\end{equation}
where $T$ denotes the scattering amplitude, $V$ denotes the interaction kernel (potential), $p'$ and $p''$
are the three-momenta of the initial and final states in the center-of-mass frame, respectively, and $z$
is the center-of-mass energy. The channel labels $\mu,\nu$ and $\kappa$ denote the meson-baryon system with
specific isospin ($I$), angular momentum ($J$, up to $9/2$), spin ($S$) and orbital angular momentum
$L$. $G_{\kappa}(p,z)$ is the propagator of the intermediate channel: 
\begin{equation}\label{Gdef}
	G_{\kappa}(z,p)=
	\big[z-E_\kappa-\omega_\kappa-\Sigma_\kappa(z,p)+i0^+\big]^{-1}\ ,
\end{equation}
where $\Sigma_\kappa$ is the self-energy of the unstable particle ($\rho$, $\sigma$ or $\Delta$) in the
effective channel $\kappa$. When $\kappa$ is not an effective channel, $\Sigma_\kappa=0$. Further, $E_\kappa,\,
\omega_\kappa$ denote the energies of the baryon and the meson in channel $\kappa$, respectively, with the
relativistic dispersion relation, e.g. $E_\kappa=\sqrt{p^2+M_\kappa^2}$. The whole formalism is established in the framework of
time-ordered perturbation theory~\cite{schweber1964}, which, together with an expansion in the partial-wave basis,
ensures the integral in Eq.~\eqref{scequ} is only one-dimensional. 

To simplify the calculation, we further perform the following separation: 
\begin{equation}\label{tsep}
	T=T^{NP}+T^{P}\ ,
\end{equation}
where $T^{P}$ and $T^{NP}$ are the pole part and the non-pole part, respectively.
The non-pole part $T^{NP}$ is generated only by the potentials without $s$-channel bare states ($V^{NP})$: 
\begin{equation}\label{tnpdef}
\begin{split}
  &T_{\mu\nu}^{NP}(p'',p',z)=V_{\mu\nu}^{NP}(p'',p',z)\\
  &+\sum_{\kappa}\int_0^\infty p^2 dpV_{\mu\kappa}^{NP}(p'',p,z)G_{\kappa}(p,z)T_{\kappa\nu}^{NP}(p,p',z)\ ,
\end{split}
\end{equation}
whereas the pole part is
\begin{equation}\label{Tpequ}
    T_{\mu\nu}^{P}(p'',p',z)=\sum_{i,j}\Gamma_{\mu,i}^a(p'')D_{ij}^{}(z)\Gamma_{\nu,j}^c(p')\ .
\end{equation}
Here, $i,j$ are the indices of the $s$-channel bare states in a given partial wave, $\Gamma_{\mu,i}^{a(c)}$ is the dressed vertex
function describing the annihilation (creation) of the $i$th state to channel $\mu$. The dressed
propagator of the $s$-channel state is denoted as $D$, which is related to the self-energies $\Sigma_{ij}$: 
\begin{equation}\label{Tpequdef}
    \begin{split}
    &\Gamma_{\mu,i}^a(p'')=\gamma_{\mu,i}^a(p'')\\
    &+\sum_{\kappa}\int_0^\infty p^2 dp T_{\mu\kappa}^{NP}(p'',p,z)G_\kappa(p,z)\gamma_{\kappa,i}^a(p)\ ,\\
    &\Gamma_{\nu,j}^c(p')=\gamma_{\nu,j}^c(p')\\
    &+\sum_{\kappa}\int_0^\infty p^2 dp \gamma_{\kappa,j}^c(p)G_\kappa(p,z)T_{\kappa\nu}^{NP}(p,p',z)\ ,\\
    &D^{-1}_{ij}(z)= \delta_{ij}(z-m_i^b)-\Sigma_{ij}(z)\ ,\\
    &\Sigma_{ij}(z)=\sum_{\kappa}\int_0^\infty p^2 dp \gamma_{\kappa,i}^c(p)G_\kappa(p,z)\Gamma_{\kappa,j}^a(p)\ .
    \end{split}
\end{equation}
Here, the $\gamma$'s are bare vertices and $m_i^b$ is the bare mass of the $i$th bare state. All the relevant
expressions can be found in Ref.~\cite{Wang:2022osj} and its supplemental material. Photoproduction reactions are described in a semi-phenomenoligical approach with $T_{\mu\nu}$ as the hadronic final-state interaction, see Ref.~\cite{Ronchen:2014cna} for details. The free parameters
in this model are determined by a global fit to almost all available data. The resonances are extracted
by scanning the complex energy plane on the unphysical Riemann sheet of the full $T$-matrix, for details see Ref.~\cite{Doring:2009yv}.
Note that the decomposition of the scattering amplitude into a pole and a non-pole part in Eq.~\eqref{tsep} is ambiguous since resonance poles can be produced not only by the $s$-channel bare states but can also be dynamically generated in the
non-pole part in Eq.~\eqref{tnpdef}. However, due to the complicated multi-channel space and strong dressing effects inherent in the model, a clear interpretation of a certain observed pole as an $s$-channel or ``genuine" state in contrast to a dynamical one is often very difficult.

\subsection{Spectral density functions of the model}
The J{\"u}lich-Bonn model can be rewritten in terms of the states in the Hilbert space. The free
continuous states are denoted as $|\mu,p\rangle$, which satisfy
\begin{equation}\label{stamupdef}
    \langle\nu,q|\mu,p\rangle=p^{-2}\delta_{\mu\nu}\delta(p-q)\ ,
\end{equation}
whereas the $s$-channel bare states are $|\Phi_n\rangle$, with 
\begin{equation}\label{stapsidef}
    \langle \Phi_m|\Phi_n\rangle=\delta_{mn}\ . 
\end{equation}
Further conditions are 
\begin{equation}\label{otho}
	\langle \Phi_n|\mu,p \rangle=\langle \mu,p |\Phi_n\rangle=0\ ,
\end{equation}
and
\begin{equation}\label{comp}
	\sum_n |\Phi_n\rangle\langle\Phi_n|+\sum_\mu \int_0^\infty p^2dp | \mu,p \rangle\langle \mu,p |=\hat{I}\ .
\end{equation}
The Hamiltonian is written as 
\begin{equation}
	\hat{H}=\hat{H}_0+\hat{H}_I\ ,
\end{equation}
with the free part 
\begin{equation}\label{H0def}
	\hat{H}_0=\sum_j m_j^b|\Phi_j\rangle\langle\Phi_j|+
	\sum_\mu \int p^2dp (E_\mu+\omega_\mu) | \mu,p \rangle\langle \mu,p |\ ,
\end{equation}
and the interaction part 
\begin{equation}\label{HIdef}
\begin{split}
	\hat{H}_I&=\sum_i\sum_\mu \int p^{\prime\prime 2}dp'' \left[\gamma_{\mu,i}^a(p'')| \mu,p'' \rangle\langle \Phi_i |
	+\text{h.c.}\right]\\
	&+\sum_{\mu,\nu}\iint dp'dp''(p'p'')^2 V_{\mu\nu}^{NP}(p'',p') | \mu,p'' \rangle\langle \nu,p' |\ .
\end{split}
\end{equation}

The scattering equation~\eqref{scequ} is fully equivalent to the eigenequation of the energy $z$: 
\begin{equation}\label{psiequ}
	(\hat{H}_0+\hat{H}_I)|z\rangle=z|z\rangle\ ,
\end{equation}
where a physical scattering state is denoted as $|z\rangle$. The expansion is very similar to Eq.~\eqref{pedasolsta2ch}: 
\begin{equation}\label{solstadef}
    |z\rangle=\sum_ic_i(z)|\Phi_i\rangle+\sum_\alpha\int k^2\,dk\,\chi_\alpha(z,k)|\alpha,k\rangle\ . 
\end{equation}
The spectral density function of the scattering state on the $i$th $s$-channel bare state can be defined as 
\begin{equation}\label{SDdef}
    w_i(z)\equiv |\langle\Phi_i|z\rangle|^2=|c_i(z)|^2\ . 
\end{equation}
The sum rule for every $i$ is
\begin{equation}\label{SDsum}
	\mathcal{Z}_{B,i}+\int_{m_\pi+m_N}^\infty dz w_i(z)=1\ ,
\end{equation}
where $\mathcal{Z}_{B,i}$ is the partial elementariness of a possible bound state on the $i$th $s$-channel
bare state. Note that in the J{\"u}lich-Bonn model the only bound state is the nucleon itself in the $P_{11}$
wave of $\pi N$ scattering. 

The J{\"u}lich-Bonn model is much more complicated than the toy model discussed previously, so the
analytical expressions cannot be solved explicitly. Nevertheless the spectral density functions in
Eq.~\eqref{SDdef} can still be obtained by the imaginary part of the propagator in Eq.~\eqref{Tpequdef}: 
\begin{equation}\label{SDSig}
    w_i(z)=-\frac{1}{\pi}\text{Im}D_{ii}(z)\ ,
\end{equation}
and the propagator can be calculated numerically. For the resonance pole at $z=M_R-{i}\Gamma_R/2$,
the elementariness corresponding to the $i$th bare state, denoted as $\mathcal{Z}_i$, is evaluated as 
\begin{equation}\label{ZReval}
    \mathcal{Z}_i \simeq \frac{\int_{M_R-\Gamma_R}^{M_R+\Gamma_R}w_i(z)dz}
    {\int_{M_R-\Gamma_R}^{M_R+\Gamma_R}BW(z)dz}\ .
\end{equation}
Note that the $\Delta E$ in Eq.~\eqref{pedaZReval} is chosen as the pole width of the resonance here. To avoid the
uncertainties of the overlap between two resonances in the same partial wave, the value of $\Delta E$ should not be too
large. At last, when there are more than one bare states in the same partial wave, the ``total elementariness''
can be estimated as
\begin{equation}\label{totalZ}
    \mathcal{Z}=1-\prod_i(1-\mathcal{Z}_i)\ .
\end{equation}

\subsection{Estimation of the model uncertainties}
We emphasize that the number of resonance poles does not correspond to the number of $s$-channel
bare states. The free parameters of the model can move the pole of an $s$-channel state to distant regions in the complex plane
beyond the reach of the pole searching, while the $T^{NP}$ part in Eq.~\eqref{tsep} can also contain
``dynamically generated poles''.
Actually, the separation of the amplitude and the number of $s$-channel states are model-dependent, which also holds for Eq.~\eqref{ZReval} that does depend on those, and hence it can only be
regarded as a naive indication from the model. 
Despite the difficulties of getting rid of these model dependences, we will try to give some rough
estimation for the uncertainties.  

\subsubsection{Locally constructed spectral density functions}
The first proposal is to locally construct (lc) a spectral density function for every individual resonance,
using only the pole positions and the residues. We can first construct the on-shell $T$ amplitude near the pole: 
\begin{equation}\label{Tcst}
	T_{\alpha\beta}^{\text{lc}}(z)=
	\frac{c g_\alpha g_\beta f_\alpha^a(q_{\alpha z})f_\beta^c(q_{\beta z})}{z-M_0-\sum_\kappa g^2_\kappa L_\kappa(z)}
    +\cdots\ ,
\end{equation}
where $q_{\alpha z}$ is the on-shell momentum for energy $z$ in channel $\alpha$, the $g$'s are real coupling
constants, $f_\alpha^{a(c)}$ is the vertex function of the resonance annihilation (creation) without the coupling
constant, $M_0$ is a mass parameter to be determined by the pole position, ``$\cdots$'' refers to the other
contributions apart from the pole, and $L_\kappa$ is the loop function of channel $\kappa$ with respect to the
vertex functions: 
\begin{equation}\label{Lcstdef}
	L_\kappa(z)\equiv \int_0^\infty p^2 dp\, G_{\kappa}^{}(p,z)f_\alpha^a(q_{\kappa z})f_\alpha^c(q_{\kappa z})\ .
\end{equation}
Eq.~\eqref{Tcst} can be regarded as a generalization of the Laurent expansion. For simplicity we just choose
the form of the vertex function $f$ the same as the bare vertices in J{\"u}lich-Bonn model (together with the
regulator, with the cut-off parameters set as the pole masses). The $\sigma N$ channel does not have bare
vertices in this model. Instead, we take $f_{\sigma N}^{a,c}(q)\sim q^l$ with $l$ the angular momentum. The
number $c$ is an extra compensation factor. Specifically, as we already have the residue $r_\alpha$ at
the pole $z=M_R-{i}\Gamma_R/2$, all the parameters can be given (here, the subscript ``$1$'' refers to
the $\pi N$ channel): 
\begin{equation}\label{Tcstsolve}
\begin{split}
	&h_\kappa\equiv\frac{g_\kappa^2}{g_1^2}=\Big|\frac{r_\kappa f_1^a}{r_1 f_\kappa^a}\Big|^2\ ,\\
	&g_1^2=-\frac{\Gamma_R}{2\sum_\kappa h_\kappa \text{Im}(L_\kappa^{II})}\ ,\\
	&M_0=M_R-g_1^2\sum_\kappa h_\kappa \text{Re}(L_\kappa^{II})\ ,\\
	&c=\frac{r_1^2}{g_1^2f_1^af_1^c}\left(1-g_1^2\sum_\kappa h_\kappa \frac{d}{dz}L_\kappa^{II}\Big|_{z=M_R-i\Gamma_R/2}\right)\ .
\end{split}
\end{equation}
This parametrization cannot reproduce the phase of the residues. However, the phases mainly affect the
interference behavior of a resonance with the others. Here we construct the amplitude only for every
individual state with the energy near the pole mass. Finally, the locally constructed spectral density function is: 
\begin{equation}\label{SDFcst}
	w^{\text{lc}}(z)=-\frac{1}{\pi}\text{Im}\left[z-M_0-\sum_\kappa g^2_\kappa L_\kappa(z)\right]^{-1}\ .
\end{equation}
According to the asymptotic behavior of $L_\kappa$ and the analytic properties, the sum rule in Eq.~\eqref{pedasr}
still holds for this construction. 

Note that Eq.~\eqref{Tcst} is still too simple to simulate the amplitude in J{\"u}lich-Bonn model. This formalism
fails if $\sum_\kappa h_\kappa \text{Im}\,L_\kappa^{II}>0$, which does not happen when the state is narrow enough.
In this case the uncertainty is of course large. We can employ a ``plan B'' and directly assign the couplings
in Eq.~\eqref{SDFcst} as the absolute values of the dimensionless normalized residues defined in
PDG~\cite{Workman:2022ynf}: 
\begin{equation}\label{Nmres}
	g_\kappa^2\to\Bigg|\sqrt{\frac{2\pi \rho_\kappa}{\Gamma_R}}r_\kappa\Bigg|^2\ ,
\end{equation}
where $\rho_\kappa={q_{\kappa z}}E_\kappa\omega_\kappa/z$ is a kinematic factor. The fixed $g_\kappa$'s do not
automatically lead to the correct pole position. In fact, we have to give $M_0$ an imaginary part compensating
for the other implicit effects. This destroys the sum rule in Eq.~\eqref{pedasr} and sometimes makes
the spectral density function negative. Before applying the plan B, one has to check if such bad
features are significant. 

\subsubsection{Complex compositenesses of the Gamow states}
The second proposal is to adopt the Gamow states and their compositenesses. Totally similar to
Eq.~\eqref{pedaXgeneral}, the compositeness for channel $\kappa$ is 
\begin{equation}\label{complX}
	X_\kappa\equiv\int_{C_+} p^2 dp\ \mathfrak{r}_\kappa^2(p) G_{\kappa}^2(p,z_{\text{pole}})\ ,
\end{equation}
where $\mathfrak{r}_\kappa$ is the off-shell residue from the J\"{u}lich-Bonn model, and $C_+$ is the
deformed contour to ensure the correct Riemann sheet. That means when the pole is lower to the threshold
of channel~$\kappa$, the integral contour is $[0,+\infty]$. When the pole is higher to the threshold,
the extra contributions from the deformed contour are estimated by the following extrapolation: 
\begin{equation}\label{resextra}
	\mathfrak{r}_\kappa(p)\to \frac{r_\kappa}{f_\kappa^a(p_{\text{pole}})}f_\kappa^a(p)\ ,
\end{equation}
with $f$  the bare vertex function just used in Eq.~\eqref{Tcst}. This extrapolation ensures that the on-shell
residue $r_\kappa$ is correctly reproduced. The complex elementariness of a state is: 
\begin{equation}
	Z=1-\sum_\kappa X_\kappa\ .
\end{equation}
At last, to make a comparison to the probabilities from the spectral density functions, we can also use the
naive measure proposed in Ref.~\cite{Sekihara:2016xnq} to get the rates between $0$ and $1$: 
\begin{equation}\label{ZXmeasure}
	\tilde{X}_\kappa\equiv\frac{|X_\kappa|}{\sum_\alpha|X_\alpha|+|Z|}\ ,\quad 
	\tilde{Z}\equiv\frac{|Z|}{\sum_\alpha|X_\alpha|+|Z|}\ .
\end{equation}

Note that the basic philosophy of this method is very different from the spectral density functions, so we do
not expect perfect matches of those results. However, it is the difference in nature that makes the adoption of
the Gamow states somehow instructive as the complex quantities only depend on the pole parameters, hence they are
free from the ambiguities of choosing the integral interval, or the overlaps of the resonances. Another
advantage is the complex compositenesses directly show which channel dominants the composition. 

\section{Numerical results}\label{sec:result}
\subsection{Selection of the resonances}
There are two recent results of the J{\"u}lich-Bonn model: the ``J\"uBo\_{}omegaN'' solution~\cite{Wang:2022osj},
which is based on the study of purely hadronic, pion-induced reactions, with the channel space extended to $\omega N$, and
the ``J\"uBo2022'' solution~\cite{Ronchen:2022hqk}, which includes also photoproduction reactions with $\gamma p\to K\Sigma$
newly considered, while the $\omega N$ channel is absent. In this work we use the output from the latter. Actually
some states are not significant in the hadronic part, e.g. $N(1900)\,\frac{3}{2}^+$, but play an important role in photoproduction. Moreover, the quality and quantity of the available photoproduction data supersede the ones of the pion-induced data by far. We therefor consider the resonance parameters extracted in J\"uBo2022 to be more reliable. Note, however, that we
cannot exclude the possibility that some of the results might change when the $\omega N$ channel (or further channels)
is (are) considered in the future.

The results for higher partial waves are always less stable~\cite{Ronchen:2022hqk}. Therefore we only study the $J\leq 5/2$ $N^*$ states and the $J\leq 3/2$ $\Delta$ states. Although there are two $\Delta$ states with
$J=5/2$ listed by the Particle Data Group~\cite{Workman:2022ynf}, the $\Delta(1905)\,\frac{5}{2}^+$ and $\Delta(1930)\,\frac{5}{2}^-$, we exclude those states since they proved to be rather unstable in recent J\"uBo analyses, c.f. the discussion in Ref.~\cite{Ronchen:2022hqk} for details. This may be related to the fact that the data base for the $I=3/2$
channels is still smaller than for the $I=1/2$
channels. 
In addition, the states with widths $\Gamma_R>300$~MeV are not considered, since the spectral density
functions does not work well for broad states. Specifically, in this model the $\Delta(1910)\,\frac{3}{2}^+$
state is always not significant and very broad ($\Gamma_R\simeq~550$ MeV). It is thus excluded. 

Moreover, we expect that the uncertainties of the resonance parameters estimated in Ref.~\cite{Ronchen:2022hqk} are negligible compared to the systematic uncertainties from the
criteria of the elementariness.
Therefore we do not use the uncertainties of the resonance parameters in the current calculations. 

In summary,  $13$ states out of $25$ in Ref.~\cite{Ronchen:2022hqk} are selected and summarized in
Tab.~\ref{tab:chsbare}. In the $P_{11}$ wave there are two bare $s$-channel  states, of which the first is  the
bare nucleon. By applying the sum rule Eq.~\eqref{SDsum}, the nucleon wave function renormalization constant 
is $Z_N\simeq 0.54$\footnote{In our model the bare mass and bare couplings of the
nucleon are always adjusted such that its physical mass is $938$~MeV. }. 
Note that we also use the bare nucleon to measure the elementarinesses of the other states in the $P_{11}$ wave.
\begin{table}[t!]
	\small
    \begin{ruledtabular}
	\begin{tabular}{cccc} 
		$I(J^P)$ & $L_{2I\ 2J}$ & $N_s$ & States to be studied\\
		\hline
		$\frac{1}{2}\Big(\frac{1}{2}^-\Big)$ & $S_{11}$ & $2$ & $N(1535)$,$N(1650)$\\
		$\frac{3}{2}\Big(\frac{1}{2}^-\Big)$ & $S_{31}$ & $1$ & $\Delta(1620)$\\
		$\frac{1}{2}\Big(\frac{1}{2}^+\Big)$ & $P_{11}$ & $2$ & $N(1440)$,$N(1710)$\\
		$\frac{1}{2}\Big(\frac{3}{2}^+\Big)$ & $P_{13}$ & $2$ & $N(1720)$,$N(1900)$\\
		$\frac{3}{2}\Big(\frac{3}{2}^+\Big)$ & $P_{33}$ & $2$ & $\Delta(1232)$,$\Delta(1600)$\\
		$\frac{1}{2}\Big(\frac{3}{2}^-\Big)$ & $D_{13}$ & $1$ & $N(1520)$\\
		$\frac{3}{2}\Big(\frac{3}{2}^-\Big)$ & $D_{33}$ & $1$ & $\Delta(1700)$\\
		$\frac{1}{2}\Big(\frac{5}{2}^-\Big)$ & $D_{15}$ & $1$ & $N(1675)$\\
		$\frac{1}{2}\Big(\frac{5}{2}^+\Big)$ & $F_{15}$ & $1$ & $N(1680)$\\
	\end{tabular}
    \end{ruledtabular}
	\caption{Partial waves, the number of $s$-channel bare states ($N_s$), and the states to be studied. The $L_{2I\ 2J}$ notation is only for the $\pi N$ channel. }
	\label{tab:chsbare}
\end{table}

At last, as mentioned in the last section, the estimations of the uncertainties require the residues as input.
The residues of the $\sigma N$ channel have not been published in Ref.~\cite{Ronchen:2022hqk}, and are
listed here in Tab.~\ref{tab:sigNres}. Most of the values are much smaller (of order $10^{-6}$~MeV$^{-1/2}$) than the residues of the other
channels (of order $10^{-3}$~MeV$^{-1/2}$), since in the J\"uBo2022 model the bare resonance vertices do not couple to $\sigma N$. The $\sigma N$ residues receive contributions from  coupled-channel and non-pole effects and are thus extremely small. 
\begin{table}[t!]
	\small
    \begin{ruledtabular}
	\begin{tabular}{cccc} 
		State & $r_{\sigma N}$ ($10^{-6}$ MeV$^{-1/2}$)\\
        \hline
		$N(1535)\,\frac{1}{2}^-$ & $0.294-0.207i$\\
		$N(1650)\,\frac{1}{2}^-$ & $-0.163+0.296i$\\
		$N(1440)\,\frac{1}{2}^+$ & $3.948-8.295i$\\
		$N(1710)\,\frac{1}{2}^+$ & $1.303-14.334i$\\
		$N(1720)\,\frac{3}{2}^+$ & $-0.564-0.212i$\\
		$N(1900)\,\frac{3}{2}^+$ & $2.707-5.367i$\\
		$N(1520)\,\frac{3}{2}^-$ & $0.935+3.844i$\\
		$N(1675)\,\frac{5}{2}^-$ & $-0.083-0.032i$\\
		$N(1680)\,\frac{5}{2}^+$ & $0.198-0.091i$\\
	\end{tabular}
    \end{ruledtabular}
	\caption{Residues of the selected $N^*$ states in the $\sigma N$ channel given by the ``J\"uBo2022'' solution. }
	\label{tab:sigNres}
\end{table}

\subsection{Analyses of the \texorpdfstring{$N^*$}{N*} states}
For each selected state, the elementarinesses from three methods will be shown, namely $\mathcal{Z}_i~(i=1,2,{\rm tot})$
from the naive indication of the model (Eq.~\eqref{SDSig}), $\mathcal{Z}^{\text{lc}}$ from the local
construction (Eq.~\eqref{SDFcst}), and the naive measure  $\tilde{Z}$ of the Gamow states (Eq.~\eqref{ZXmeasure}).
The first is further labelled by subscripts: $\mathcal{Z}_1(\mathcal{Z}_2)$ for the partial elementariness on the first
(second) bare $s$-channel state, and $\mathcal{Z}_{\rm tot}$ for the ``total'' elementariness in Eq.~\eqref{totalZ}. Note that in a complex coupled-channel environment with strong dressing effects as in the J\"uBo model, in cases with two states per partial wave it is often difficult to identify a pole as unambiguously induced by a specific bare $s$-channel state. Therefore we list 
$\mathcal{Z}_1$ and $\mathcal{Z}_2$ for each pole.

\begin{figure*}[t!]
	\centering
	\includegraphics[width=0.75\textwidth]{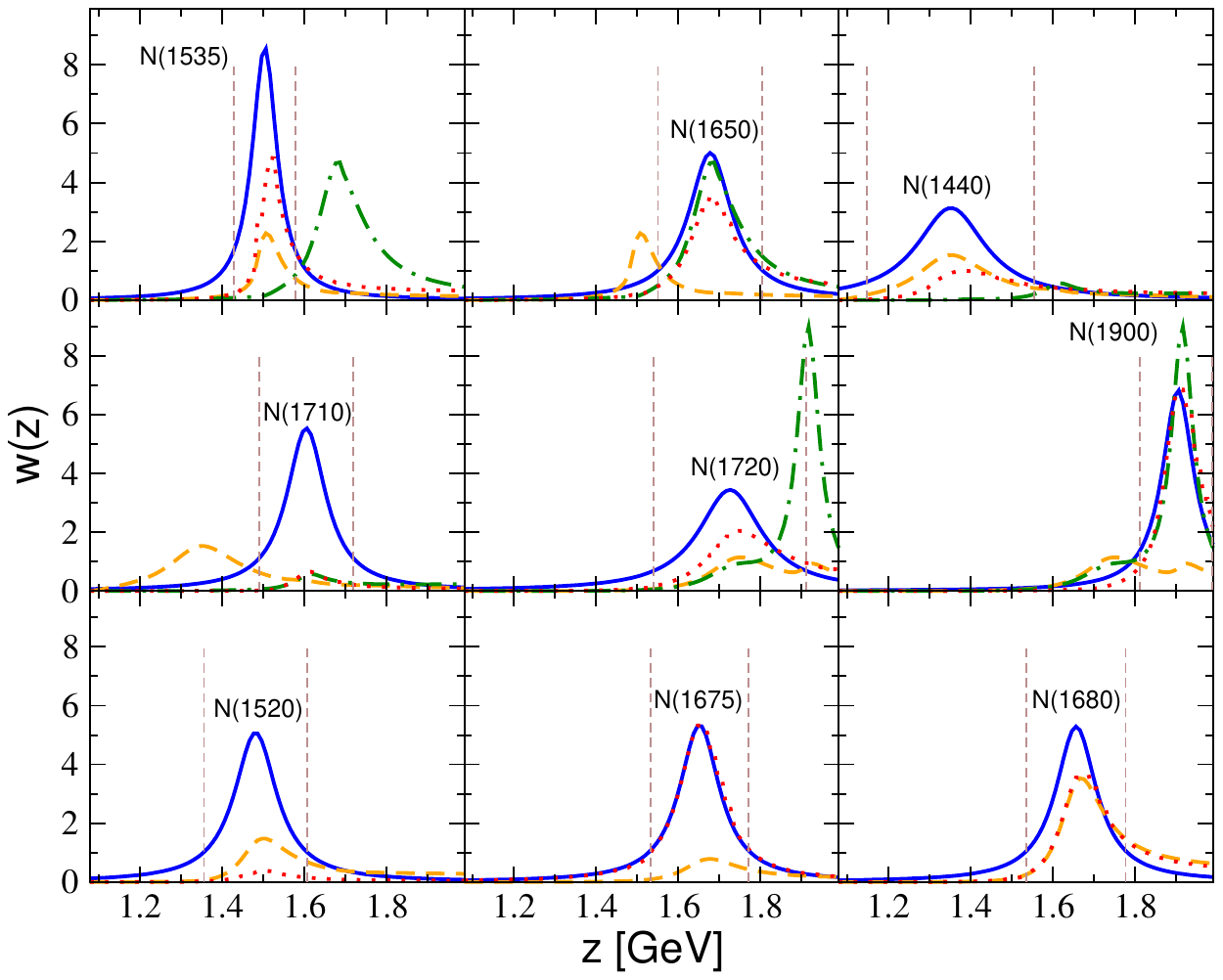}
	\caption{The spectral density functions for all the selected $N^*$ states. Blue solid line:
          the Breit-Wigner denominator in Eq.~\eqref{pedabww}. Orange dashed (green dash-dotted) line: the
          1st (2nd) spectral density function from the model. Red dotted line: the locally constructed
          function in Eq.~\eqref{SDFcst} (for $N^*(1675)$ Eq.~\eqref{Nmres} is applied). The vertical lines label the integral region of Eq.~\eqref{ZReval}. }
	\label{fig:Nsdf}
\end{figure*}

The elementarinesses of the selected $N^*$ states are summarized in Tab.~\ref{tab:selNresult}. The complex
quantities of the Gamow states are listed in Tabs.~\ref{tab:selNX2} and \ref{tab:selNX3}. Since the
$\sigma N$ residues are always very small, we skip the discussion of the compositenesses in the $\sigma N$ channel.
The spectral density functions are plotted in Fig.~\ref{fig:Nsdf}. The results are explained and discussed one
by one in what follows. 
\begin{table*}[t!]
	\footnotesize
    \begin{ruledtabular}
	\begin{tabular}{ccccccc} 
		State & Pole position (MeV) & $\mathcal{Z}_1$ & $\mathcal{Z}_2$ & $\mathcal{Z}_{\rm tot}$ & $\mathcal{Z}^{\text{lc}}$ & $\tilde{Z}$\\
        \hline
		$N(1535)\,\frac{1}{2}^-$ & $1504-37i$ & $24.8\%$ & $5.6\%$ & $29.0\%$ & $50.8\%$ & $39.4\%$\\
		$N(1650)\,\frac{1}{2}^-$ & $1678-64i$ & $13.4\%$ & $91.7\%$ & $92.8\%$ & $70.5\%$ & $8.5\%$\\
		$N(1440)\,\frac{1}{2}^+$ & $1353-102i$ & $48.7\%$ & $1.7\%$ & $49.5\%$ & $31.5\%$ & $36.9\%$\\
		$N(1710)\,\frac{1}{2}^+$ & $1605-58i$ & $11.5\%$ & $10.3\%$ & $20.6\%$ & $10.2\%$ & $40.3\%$\\
		$N(1720)\,\frac{3}{2}^+$ & $1726-93i$ & $34.1\%$ & $68.5\%$ & $79.3\%$ & $62.5\%$ & $41.4\%$\\
		$N(1900)\,\frac{3}{2}^+$ & $1905-47i$ & $19.9\%$ & $100\%$ & $100\%$ & $99.9\%$ & $38.5\%$\\
		$N(1520)\,\frac{3}{2}^-$ & $1482-63i$ & $29.4\%$ & $\cdots$ & $29.4\%$ & $7.2\%$ & $40.4\%$\\
		$N(1675)\,\frac{5}{2}^-$ & $1652-60i$ & $16.6\%$ & $\cdots$ & $16.6\%$ & $100\%$ (F) & $61.8\%$\\
		$N(1680)\,\frac{5}{2}^+$ & $1657-60i$ & $67.9\%$ & $\cdots$ & $67.9\%$ & $69.9\%$ & $55.0\%$\\
	\end{tabular}
    \end{ruledtabular}
    \caption{The elementarinesses of the selected $N^*$ states. The label ``(F)'' means that the
      local construction of Eq.~\eqref{Tcstsolve} has failed and Eq.~\eqref{Nmres} is used instead. }
	\label{tab:selNresult}
\end{table*}
\begin{table*}[t!]
	\footnotesize
    \begin{ruledtabular}
	\begin{tabular}{cccccc} 
		State & $X_{\pi N}$ & $X_{\eta N}$ & $X_{K\Lambda}$ & $X_{K\Sigma}$ & $Z$\\
        \hline
		$N(1535)\,\frac{1}{2}^-$ & \St{$0.12+0.14i$\\$(7.5\%)$} & 
        \St{$0.67+0.57i$\\$(35.8\%)$} & 
        \St{$0.03-0.01i$\\$(1.3\%)$} & \St{$0.01-0.03i$\\$(1.2\%)$} & \St{$0.32-0.92i$\\$(39.4\%)$} \\
        \hline
		$N(1650)\,\frac{1}{2}^-$ & \St{$0.17+0.28i$\\$(22.2\%)$} & 
        \St{$0.60-0.15i$\\$(41.4\%)$} & 
        \St{$0.03+0.03i$\\$(3.0\%)$} & \St{$0.15-0.03i$\\$(10.2\%)$} & \St{$0.12+0.02i$\\$(8.5\%)$} \\
        \hline
		$N(1440)\,\frac{1}{2}^+$ & \St{$0.69+0.37i$\\$(59.0\%)$} & 
        \St{$0.00+0.00i$\\$(0.2\%)$} & 
        \St{$0.00-0.00i$\\$(0.0\%)$} & \St{$0.00+0.00i$\\$(0.0\%)$} & \St{$0.34-0.36i$\\$(36.9\%)$} \\
        \hline
		$N(1710)\,\frac{1}{2}^+$ & \St{$0.04-0.00i$\\$(1.4\%)$} & 
        \St{$0.87+0.97i$\\$(44.9\%)$} & 
        \St{$-0.10+0.18i$\\$(6.9\%)$} & \St{$-0.00+0.01i$\\$(0.3\%)$} & \St{$0.02-1.17i$\\$(40.3\%)$} \\
        \hline
		$N(1720)\,\frac{3}{2}^+$ & \St{$-1.16-3.01i$\\$(32.6\%)$} & 
        \St{$0.04-0.04i$\\$(0.5\%)$} & 
        \St{$0.03+0.05i$\\$(0.6\%)$} & \St{$-0.08+0.13i$\\$(1.5\%)$} & \St{$2.71+3.09i$\\$(41.4\%)$} \\
        \hline
		$N(1900)\,\frac{3}{2}^+$ & \St{$-0.00+0.00i$\\$(0.1\%)$} & 
        \St{$-0.02+0.00i$\\$(0.2\%)$} & 
        \St{$0.39+0.33i$\\$(6.0\%)$} & \St{$0.90-0.17i$\\$(10.7\%)$} & \St{$-3.04+1.22i$\\$(38.5\%)$} \\
        \hline
		$N(1520)\,\frac{3}{2}^-$ & \St{$0.19+0.39i$\\$(15.9\%)$} & 
        \St{$-0.00+0.00i$\\$(0.0\%)$} & 
        \St{$0.00+0.00i$\\$(0.0\%)$} & \St{$0.00-0.00i$\\$(0.0\%)$} & \St{$0.87-0.66i$\\$(40.4\%)$} \\
        \hline
		$N(1675)\,\frac{5}{2}^-$ & \St{$0.02+0.17i$\\$(9.4\%)$} & 
        \St{$0.02-0.06i$\\$(3.3\%)$} & 
        \St{$-0.00+0.00i$\\$(0.0\%)$} & \St{$0.00+0.00i$\\$(0.0\%)$} & \St{$1.01-0.49i$\\$(61.8\%)$} \\
        \hline
		$N(1680)\,\frac{5}{2}^+$ & \St{$0.22+0.50i$\\$(36.3\%)$} & 
        \St{$-0.00-0.00i$\\$(0.0\%)$} & 
        \St{$0.00-0.00i$\\$(0.0\%)$} & \St{$0.00-0.00i$\\$(0.0\%)$} & \St{$0.68-0.48i$\\$(55.0\%)$} \\
	\end{tabular}
    \end{ruledtabular}
    \caption{The compositenesses and elementarinesses of the selected $N^*$ Gamow states (two-body channels).
      The percentages in the brackets are the naive measures from Eq.~\eqref{ZXmeasure}. }
	\label{tab:selNX2}
\end{table*}
\begin{table*}[t!]
	\footnotesize
    \begin{ruledtabular}
	\begin{tabular}{cccccc} 
		State & $X_{\rho N}(1)$ & $X_{\rho N}(2)$ & $X_{\rho N}(3)$ & $X_{\pi\Delta}(1)$ & $X_{\pi\Delta}(2)$\\
        \hline
		$N(1535)\,\frac{1}{2}^-$ & 
        \St{$-0.17+0.27i$\\$(13.0\%)$} & $\cdots$ & \St{$-0.00+0.00i$\\$(0.2\%)$} & 
        $\cdots$ & \St{$0.02-0.04i$\\$(1.6\%)$} \\
        \hline
		$N(1650)\,\frac{1}{2}^-$ & 
        \St{$-0.02-0.11i$\\$(7.4\%)$} & $\cdots$ & \St{$-0.01-0.06i$\\$(4.3\%)$} & 
        $\cdots$ & \St{$-0.04+0.02i$\\$(3.0\%)$} \\
        \hline
		$N(1440)\,\frac{1}{2}^+$ &
        \St{$-0.00+0.01i$\\$(0.6\%)$} & \St{$-0.01+0.00i$\\$(0.6\%)$} & $\cdots$ & 
        \St{$-0.02-0.03i$\\$(2.7\%)$} & $\cdots$ \\
        \hline
		$N(1710)\,\frac{1}{2}^+$ & 
        \St{$0.00-0.00i$\\$(0.0\%)$} & \St{$-0.01+0.01i$\\$(0.3\%)$} & $\cdots$ & 
        \St{$0.17+0.01i$\\$(5.9\%)$} & $\cdots$ \\
        \hline
		$N(1720)\,\frac{3}{2}^+$ & 
        \St{$0.19-0.07i$\\$(2.1\%)$} & \St{$0.58-0.07i$\\$(5.9\%)$} & \St{$0.10-0.07i$\\$(1.2\%)$} & 
        \St{$-1.40+0.01i$\\$(14.1\%)$} & \St{$-0.01-0.01i$\\$(0.1\%)$} \\
        \hline
		$N(1900)\,\frac{3}{2}^+$ & 
        \St{$0.01-0.01i$\\$(0.1\%)$} & \St{$3.04-1.60i$\\$(40.3\%)$} & \St{$-0.00-0.00i$\\$(0.0\%)$} & 
        \St{$-0.27+0.22i$\\$(4.1\%)$} & \St{$-0.00-0.00i$\\$(0.0\%)$} \\
        \hline
		$N(1520)\,\frac{3}{2}^-$ & 
        \St{$0.00+0.00i$\\$(0.1\%)$} & \St{$-0.01+0.01i$\\$(0.5\%)$} & \St{$-0.58+0.26i$\\$(23.5\%)$} & 
        \St{$0.01-0.00i$\\$(0.3\%)$} & \St{$0.52-0.00i$\\$(19.3\%)$} \\
        \hline
		$N(1675)\,\frac{5}{2}^-$ & 
        \St{$-0.00-0.00i$\\$(0.1\%)$} & \St{$-0.14+0.26i$\\$(16.2\%)$} & \St{$0.00-0.01i$\\$(0.4\%)$} & 
        \St{$0.10+0.12i$\\$(8.8\%)$} & \St{$-0.00+0.00i$\\$(0.0\%)$} \\
        \hline
		$N(1680)\,\frac{5}{2}^+$ & 
        \St{$0.00-0.01i$\\$(0.4\%)$} & \St{$0.00-0.00i$\\$(0.2\%)$} & \St{$-0.01-0.00i$\\$(0.7\%)$} & 
        \St{$0.00+0.00i$\\$(0.0\%)$} & \St{$0.11-0.01i$\\$(7.4\%)$} \\
	\end{tabular}
    \end{ruledtabular}
    \caption{The compositenesses and elementarinesses of the selected $N^*$ Gamow states (three-body channels).
      The percentages in the brackets are the naive measures from Eq.~\eqref{ZXmeasure}. The meaning of the channel
      indices are: $\rho N(1)$ $\to$ $|J-L|=\frac{1}{2},S=\frac{1}{2}$; $\rho N(2)$ $\to$ $|J-L|=\frac{1}{2},S=
      \frac{3}{2}$; $\rho N(3)$ $\to$ $|J-L|=\frac{3}{2},S=\frac{3}{2}$; $\pi \Delta(1)$ $\to$ $|J-L|=\frac{1}{2}$;
      $\pi \Delta(2)$ $\to$ $|J-L|=\frac{3}{2}$.}
	\label{tab:selNX3}
\end{table*}

\subsubsection*{\texorpdfstring{$N(1535)\,\frac{1}{2}^-$}{}}
As shown in Tab.~\ref{tab:selNresult}, all results from the three different methods indicate the elementariness of $N^*(1535)$ is not
large. As shown in Fig.~\ref{fig:Nsdf}, the peak structure near the pole position only shows up in the first
spectral density function, which means that the $N^*(1535)$ pole is highly related to the first $s$-channel state
in this model, and the mixture between the two bare states is weak. The locally constructed spectral density
function is bigger than  that of the model. However, it is still significantly smaller than the Breit-Wigner
peak. Note that the integral intervals $[M_R-\Gamma_R,M_R+\Gamma_R]$ for $N^*(1535)$ and $N^*(1650)$ have an overlap,
which increases the uncertainties. On the other side, Tab.~\ref{tab:selNX2} shows a large complex
compositeness of the Gamow state in the $\eta N$ channel, with the naive measure of $35.8\%$ for the composition.
It is suggested in this work that the $N^*(1535)$ tends to be composite. Actually in some other approaches the
$N^*(1535)$ resonance can be dynamically generated, see e.g.
Refs.~\cite{Kaiser:1995cy,Nieves:2001wt,Lutz:2001mi,Bruns:2010sv}. 

In fact, in the J\"uBo\_{}omegaN solution~\cite{Wang:2022osj} the coupling of the $N^*(1535)$ to the $\omega N$
channel is rather large, even though the pole position is not close to the $\omega N$ threshold. This large
coupling, as well as the $\omega N$ composition, will be investigated in the future when studying  $\omega N$ photoproduction.

\subsubsection*{\texorpdfstring{$N(1650)\,\frac{1}{2}^-$}{}}
Just like the $N^*(1535)$, in some models, e.g. Refs~\cite{Nieves:2001wt,Lutz:2001mi,Bruns:2010sv}, the
$N^*(1650)$ can also be dynamically generated. But here in our model it seems to be highly related to
the second bare state. The spectral density function of this state, no matter if directly from the model or
locally constructed, suggests a high elementariness, see Fig.~\ref{fig:Nsdf}. 

However, the compositenesses of the Gamow state leads to another result. In J\"uBo2022 the on-shell residues are
$r_{\pi N}=(9.31-0.90i)\times 10^{-3}$~MeV$^{-1/2}$ and $r_{\eta N}=(0.25-3.97i)\times 10^{-3}$~MeV$^{-1/2}$, in contrast 
the off-shell residue of the $\eta N$ channel is larger than that of $\pi N$, so the result $X_{\eta N}$ is also
large in Tab.~\ref{tab:selNX2}. Consequently the elementariness is small, with the naive measure of only $8.5\%$.
We should emphasize again that the naive measure is not mathematically a probability, it is understandable
that the naive measure does not match the probability given by the spectral density function. Nevertheless, this may
be an indication of the involved dynamics. Since the two results from the spectral density functions strongly suggest
 the elementary interpretation, we might claim the $N^*(1650)$ is possibly compact, but with visible
model uncertainties. Note that the results from the spectral density functions are compatible with the
conclusion in Ref.~\cite{Doring:2009uc}: it is necessary to include an $s$-channel bare state to generate
the $N^*(1650)$, but not necessary for $N^*(1535)$.

\subsubsection*{\texorpdfstring{$N(1440)\,\frac{1}{2}^+$}{}}
The $N^*(1440)$ is very interesting in this model, it is definitely dynamically generated in the $T^{NP}$
part~\cite{Krehl:1999km}. However, the peak structure still shows up significantly in the first spectral
density function, see Fig.~\ref{fig:Nsdf}. Meanwhile, the corresponding elementariness is moderate ($49.5\%$).
Note that the first bare state in this channel is just the bare nucleon. As already mentioned, the physical
nucleon in this model carries approximately $54\%$ composition of the bare nucleon. Here in the energy region
of $N^*(1440)$, i.e. $z\in[M_R-\Gamma_R,M_R+\Gamma_R]$, the set of scattering states carries $34\%$. The
result $49.5\%$ is obtained by further corrections of the Breit-Wigner peak in Eq.~\eqref{ZReval}, and the
sum rule Eq.~\eqref{SDsum} is not violated. 

The locally constructed spectral density function gives a smaller elementariness ($31.5\%$).
In Tab.~\ref{tab:selNX2}, the complex compositeness of the $\pi N$ channel is rather large, and the naive
measure of the elementariness is also small ($36.9\%$). However, we should admit that which channel has the
biggest composition is model-dependent. It has been indicated in Ref.~\cite{Sekihara:2021eah} that the $N^*(1440)$
has a large component of $\sigma N$ and also in the J\"uBo model the $\sigma N$ channel plays an important role in the $P_{11}$ partial wave~\cite{Ronchen:2012eg}. All in all, the $N^*(1440)$ tends to be a molecular state here.
Actually, in the studies of the quark model, the $N^*(1440)$ deviates from the simple picture of a $qqq$
state. For example in Ref.~\cite{Segovia:2015hra} the $N^*(1440)$ is depicted by a core of three valence
quarks complemented by a meson cloud, while in Ref.~\cite{Wang:2017cfp} the coupling of the $N^*(1440)$
to the $\pi N$ channel differs a lot from the prediction of the simple quark model. Note also that the Roper resonance
is often considered as breathing mode of the nucleon, see e.g. Refs.~\cite{Meissner:1984un,Zahed:1984qv}
and references therein\footnote{Specifically, the calculation of the decay width in Ref.~\cite{Meissner:1984un} does not support the interpretation of the $N^*(1440)$ state as the radial excitation of the nucleon. }. We point out that the interplay of a bare Roper state with pion loops was already considered in Ref.~\cite{Meissner:1999vr} in the framework of chiral dynamics with explicit resonance fields.

\subsubsection*{\texorpdfstring{$N(1710)\,\frac{1}{2}^+$}{}}
The quantum numbers of the $N^*(1710)$ are the same as for the $N^*(1440)$, and their elementarinesses are both not
large from the spectral density functions. As seen from Fig.~\ref{fig:Nsdf}, the first function from
the model does not show a significant structure near its pole mass, while the second has a very weak
structure. The local construction is very similar to the latter. The total elementariness of this state
does not exceed $21\%$. The complex compositeness of the $\eta N$ channel is quite large. As given in
Tab.~\ref{tab:selNX2}, the naive measure of the $\eta N$ compositeness is $44.9\%$, whereas the elementariness
is around $40.3\%$. So all the three results suggest the $N^*(1710)$ to be more of composite nature. This corroborates the findings of Ref.~\cite{Ronchen:2022hqk} that the $N^*(1710)$ is dynamically generated in the J\"uBo2022 analysis while the second bare $s$-channel of the $P_{11}$ wave pole moved far into the complex plane ($z_0=1513-i405$~MeV).

Note that the pole of $N^*(1710)$ in the J\"uBo\_{}omegaN solution~\cite{Wang:2022osj} (fit~A) is only $20$~MeV
below the $\omega N$ threshold, and the coupling is large. One may expect a quantitative change of the result
for this state with a large component in the $\omega N$ channel. However, a systematic study can only be carried
out after  $\omega N$ photoproduction is fully analysed.

\subsubsection*{\texorpdfstring{$N(1720)\,\frac{3}{2}^+$}{}}
In the $P_{13}$ wave of the $\pi N$ channel, two bare $s$-channel states are included and we observe two resonance poles, the $N^*(1720)$ and the $N^*(1900)$.
The peak of $N^*(1900)$ is very strong, which affects the analyses of the $N^*(1720)$ since it is not very narrow, c.f. Tab.~\ref{tab:selNresult}. Such interference severely increases the ambiguity. As seen from
Tab.~\ref{tab:selNresult}, the elementariness from the first spectral density function is around $34.1\%$, while the
second gives $68.5\%$, which mainly comes from the shoulder of the $N^*(1900)$ peak, see Fig.~\ref{fig:Nsdf}.
Therefore an unambiguous interpretation of the numbers above is difficult. The locally
constructed function gives a value of more than $60\%$, preferring the elementary nature. On the other hand,
the off-shell residues of the $\pi N$ and $\pi\Delta$ channels are extremely large, so the naive measure
of the elementariness is less than $50\%$. In a word, we cannot draw any firm conclusion on this state. 

\subsubsection*{\texorpdfstring{$N(1900)\,\frac{3}{2}^+$}{}}
The $N^*(1900)$ can hardly be seen in the purely hadronic observations, but is believed to be
very important for $KY$
photoproduction~\cite{Cao:2013psa,Anisovich:2017bsk,Mart:2017mwj,Clymton:2021wof,Wei:2022nqp}. In the J\"uBo model,
it is not significant if photoproduction is not considered. For instance, in the J\"uBo\_omegaN
solution~\cite{Wang:2022osj}, the pole lies in a rather distant region in the complex plane (the pole
width $\Gamma_R>600$~MeV) and the couplings are very weak. However, in the
J\"uBo2022 solution~\cite{Ronchen:2022hqk} it becomes a significant narrow resonance. 

We meet a situation similar to the $N^*(1650)$ for this state. The second spectral density function from
the model in Fig.~\ref{fig:Nsdf} simply suggests $100\%$ elementariness for this state and also
the local construction indicates nearly $100\%$ elementariness. In contrast, the complex compositeness in
Tab.~\ref{tab:selNX3} show a very large $\rho N$ composition of this state. Note that  $\rho N$ is an
effective three-body channel, for which the constraints from  data are relatively weak.

\subsubsection*{\texorpdfstring{$N(1520)\,\frac{3}{2}^-$}{}}
The $D_{13}$ wave of $\pi N$ is rather clear, there is only one $s$-channel bare state and only one resonance pole.
The spectral density functions in Fig.~\ref{fig:Nsdf}, no matter directly from the model or locally
constructed, indicate a small elementariness of this state. Furthermore, in Tab.~\ref{tab:selNX3}, the
compositions of $N^*(1520)$ in the three-body channels are relatively large. The model uncertainty for this
state seems to be small. All this suggests the $N^*(1520)$ to be composite. 

\subsubsection*{\texorpdfstring{$N(1675)\,\frac{5}{2}^-$}{}}
As shown in Fig.~\ref{fig:Nsdf}, the spectral density function directly given by the model provides a
very small elementariness for $N^*(1675)$. However, the local construction from Eq.~\eqref{Tcstsolve} fails for
this state. The residue of the $\rho N(2)$ channel, i.e. $S=3/2$ and $|J-L|=1/2$, is quite large, and the imaginary part of the $\rho N(2)$
loop function is positive in Eq.~\eqref{Tcstsolve}. We have checked the unitarity condition: every $L_\kappa^{II}$ must
have a negative imaginary part when the energy is $z-i0^+$, with $z$ real and higher than the corresponding
threshold. Even though $\text{Im}\,L_{\rho N(2)}^{II}(M_R-i0^+)<0$, at the pole $\text{Im}\,L_{\rho N(2)}^{II}(M_R-
i\Gamma_R/2)>0$. The $N^*(1675)$ is not broad, but the sign is still changed by the finite width. 
If one uses Eq.~\eqref{Nmres}, the elementariness would be nearly $100\%$, which is not so trustworthy
because the couplings are rather small, and the pole width almost totally comes from the constant width in
the mass parameter $M_0$. On the bright side,  the constant width does not cause negative values near
the pole mass, see Fig.~\ref{fig:Nsdf}. In addition, the violation of the sum rule is not large for
this construction: $\int_{m_\pi+m_N}^{\infty}w^{\text{lc}}(z)dz\simeq 0.97$, deviating from the standard value of
$1$ only by $3\%$, and this spectral density function from plan B does not show any negative value in the energy region we study.  

Furthermore, the complex compositenesses in Tabs.~\ref{tab:selNX2} and \ref{tab:selNX3} prefer a larger
elementariness, which is opposite to the spectral density function from the model. So we cannot draw a
certain conclusion on the $N^*(1675)$.

\subsubsection*{\texorpdfstring{$N(1680)\,\frac{5}{2}^+$}{}}
Unlike the $N^*(1675)$, the uncertainties of the $N^*(1680)$ in this study are not large. The spectral
density functions given by the model and the local construction agree with each other well in
Fig.~\ref{fig:Nsdf}, both of which lead to an elementariness larger than $67\%$. The complex compositenesses
in Tabs.~\ref{tab:selNX2} and \ref{tab:selNX3} may indicate non-negligible compositions of $\pi N$ and
$\pi\Delta$ channels, but the naive measure of the elementariness is still more than $55\%$. It is
expected that the $N^*(1680)$ tends to be elementary. 

Note again that the pole position of the $N^*(1680)$ is closer to the $\omega N$ threshold in the J\"uBo\_{}omegaN
solution~\cite{Wang:2022osj}, and the coupling is not small. There is a possibility that the $N^*(1680)$ becomes
more composite when $\omega N$ is considered. 
\subsection{Analyses of the \texorpdfstring{$\Delta$}{Delta} states}
\begin{figure*}[t!]
	\centering
	\includegraphics[width=0.6\textwidth]{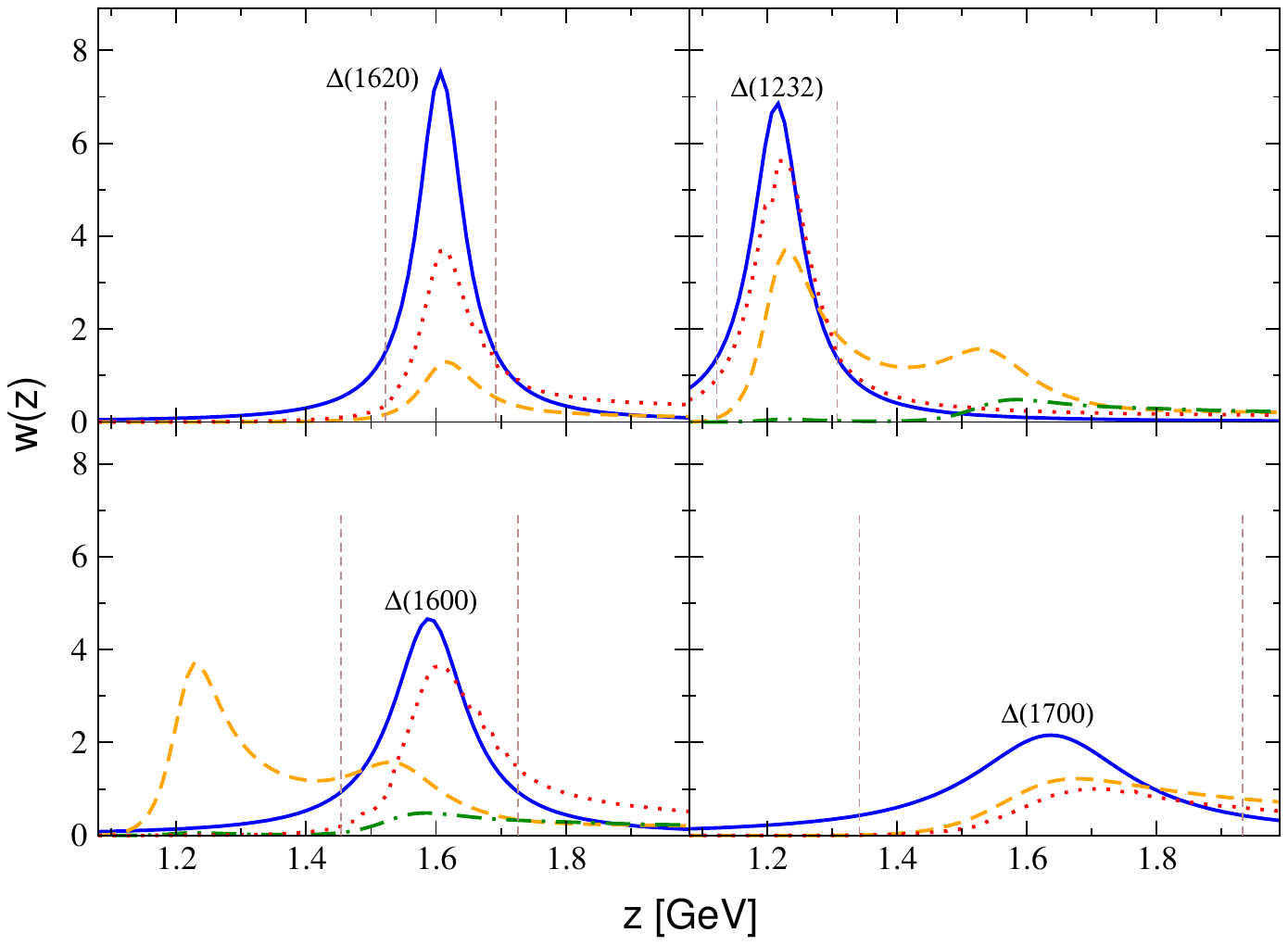}
	\caption{The spectral density functions for all the selected $\Delta$ states. Blue solid line: the Breit-Wigner
          denominator in Eq.~\eqref{pedabww}. Orange dashed (green dash-dotted) line: the 1st (2nd) spectral
          density function from the model. Red dotted line: the locally constructed function in
          Eq.~\eqref{SDFcst} (for $\Delta(1232)$ Eq.~\eqref{Nmres} is applied). The vertical lines label the integral region of Eq.~\eqref{ZReval}. }
	\label{fig:Dsdf}
\end{figure*}

The elementariness for the selected $\Delta$ states is summarized in Tab.~\ref{tab:selDresult}.
The complex quantities of the Gamow states are listed in Tab.~\ref{tab:selDX}. The corresponding
spectral density functions are plotted in Fig.~\ref{fig:Dsdf}. The results are explained and discussed
one by one in what follows. 
\begin{table*}[t!]
	\footnotesize
    \begin{ruledtabular}
	\begin{tabular}{ccccccc} 
		State & Pole position (MeV) & $\mathcal{Z}_1$ & $\mathcal{Z}_2$ & $\mathcal{Z}_{tot}$ & $\mathcal{Z}^{\text{lc}}$ & $\tilde{Z}$\\
        \hline
		$\Delta(1620)\,\frac{1}{2}^-$ & $1607-42i$ & $18.9\%$ & $\cdots$ & $18.9\%$ & $50.0\%$ & $69.4\%$\\
		$\Delta(1232)\,\frac{3}{2}^+$ & $1215-46i$ & $53.4\%$ & $0.9\%$ & $53.8\%$ & $82.6\%$ (F) & $30.5\%$\\
		$\Delta(1600)\,\frac{3}{2}^+$ & $1590-68i$ & $39.7\%$ & $13.3\%$ & $47.8\%$ & $77.5\%$ & $69.7\%$\\
		$\Delta(1700)\,\frac{3}{2}^-$ & $1637-148i$ & $59.7\%$ & $\cdots$ & $59.7\%$ & $44.9\%$ & $47.8\%$\\
	\end{tabular}
    \end{ruledtabular}
    \caption{The elementariness of the selected $N^*$ states. The label ``(F)'' means that the local
      construction of Eq.~\eqref{Tcstsolve} has failed and Eq.~\eqref{Nmres} is used instead. }
	\label{tab:selDresult}
\end{table*}
\begin{table*}[t!]
	\scriptsize
    \begin{ruledtabular}
	\begin{tabular}{ccccccccc} 
		State & $X_{\pi N}$ & $X_{\rho N}(1)$ & $X_{\rho N}(2)$ & $X_{\rho N}(3)$ & $X_{\pi\Delta}(1)$ & $X_{\pi\Delta}(2)$
         & $X_{K\Sigma}$ & $Z$\\
        \hline
		$\Delta(1620)\,\frac{1}{2}^-$ & \St{$0.10-0.03i$\\$(8.6\%)$} & 
        \St{$-0.02-0.09i$\\$(8.0\%)$} & $\cdots$ & \St{$-0.00-0.00i$\\$(0.3\%)$} & 
        $\cdots$ & \St{$0.13-0.09i$\\$(13.0\%)$} & 
        \St{$0.01+0.00i$\\$(0.7\%)$} & \St{$0.79+0.20i$\\$(69.4\%)$}\\
        \hline
		$\Delta(1232)\,\frac{3}{2}^+$ & \St{$0.63+1.16i$\\$(28.5\%)$} & 
        \St{$0.02-0.01i$\\$(0.4\%)$} & \St{$-0.00+0.01i$\\$(0.2\%)$} & \St{$0.06-0.03i$\\$(1.4\%)$} & 
        \St{$1.54-0.58i$\\$(35.4\%)$} & \St{$-0.01-0.01i$\\$(0.3\%)$} & 
        \St{$-0.00+0.15i$\\$(3.3\%)$} & \St{$-1.24-0.70i$\\$(30.5\%)$}\\
        \hline
		$\Delta(1600)\,\frac{3}{2}^+$ & \St{$-0.04+0.07i$\\$(3.5\%)$} & 
        \St{$0.00-0.01i$\\$(0.3\%)$} & \St{$0.01-0.03i$\\$(1.3\%)$} & \St{$-0.00-0.00i$\\$(0.1\%)$} & 
        \St{$-0.42+0.21i$\\$(21.0\%)$} & \St{$-0.00-0.00i$\\$(0.0\%)$} & 
        \St{$-0.08+0.05i$\\$(4.1\%)$} & \St{$1.53-0.29i$\\$(69.7\%)$}\\
        \hline
		$\Delta(1700)\,\frac{3}{2}^-$ & \St{$-0.03+0.05i$\\$(2.3\%)$} & 
        \St{$-0.00+0.01i$\\$(0.4\%)$} & \St{$-0.03+0.02i$\\$(1.3\%)$} & \St{$-0.03+0.00i$\\$(1.3\%)$} & 
        \St{$-0.01-0.03i$\\$(1.5\%)$} & \St{$0.45-0.95i$\\$(45.4\%)$} & 
        \St{$0.00+0.00i$\\$(0.0\%)$} & \St{$0.64+0.91i$\\$(47.8\%)$}\\
	\end{tabular}
    \end{ruledtabular}
	\caption{The compositeness and elementariness of the selected $\Delta$ Gamow states. The percentages in the brackets are the naive measures from Eq.~\eqref{ZXmeasure}. The meaning of the channel indices are: $\rho N(1)$ $\to$ $|J-L|=\frac{1}{2},S=\frac{1}{2}$; $\rho N(2)$ $\to$ $|J-L|=\frac{1}{2},S=\frac{3}{2}$; $\rho N(3)$ $\to$ $|J-L|=\frac{3}{2},S=\frac{3}{2}$; $\pi \Delta(1)$ $\to$ $|J-L|=\frac{1}{2}$; $\pi \Delta(2)$ $\to$ $|J-L|=\frac{3}{2}$.}
	\label{tab:selDX}
\end{table*}

\subsubsection*{\texorpdfstring{$\Delta(1620)\,\frac{1}{2}^-$}{}}
There are significant uncertainties for this state. The spectral density function given by the model
in Fig.~\ref{fig:Dsdf} is very small, which indicates strongly that this state only very weakly related to the
$s$-channel bare state. The local construction gives an exactly moderate elementariness $50.0\%$.
However, the complex compositenesses in Tab.~\ref{tab:selDX} are small, and the naive measure of the
elementariness is even $69.4\%$. Since the three scenarios lead to three different results, it is hard
to draw a conclusion for this state. Note again that although in J\"uBo2022 with the inclusion of the mixed-isospin $\gamma p\to K\Sigma$ reactions more experimental information on the $\Delta$ states was taken into account, the data base for the $\Delta$ states is still 
smaller than that for the $N^*$ states. Thus it is understandable that some outputs are uncertain. 

\subsubsection*{\texorpdfstring{$\Delta(1232)\,\frac{3}{2}^+$}{}}
As the lowest $\Delta$ state, there has been a lot of discussions on the $\Delta(1232)$ in the literature.
In the quark model the $\Delta(1232)$ can be described by a $qqq$ core, though with the Goldstone mesons as also
the degrees of freedom~\cite{Glozman:1996wq}. Ref.~\cite{Wang:2017cfp} also claims the deficiency of the
simple quark model when studying the $\Delta(1232)$. As for the studies of the compositeness, there are 
different proposals, for example Ref.~\cite{Aceti:2014ala} finds a sizeable $\pi N$ component in the resonance, 
while Ref.~\cite{Sekihara:2021eah} obtains a larger elementariness. 
Note that the interplay of the compact (bare) state with the pion loops around the nucleon was  already discussed in Ref.~\cite{Meissner:1999vr} within baryon chiral perturbation theory with explicit resonance fields, and it was shown that
both components are required to achieve a proper description of the $P_{33}$ partial wave of pion-nucleon scattering.

In the J\"uBo model the situation is even more complicated, and unfortunately we cannot draw firm conclusions. The
$\Delta$ particle appears as ground state in the initial or final $\pi\Delta$ channel and also as  $s$-channel intermediate states in various processes. In addition, there are some $u$-channel potentials with $\Delta$ being exchanged. Those three
$\Delta$'s should be physically the same, but due to practical reasons there are technical simplifications. The initial/final state
$\Delta$ only couples to the $\pi N$ channel, and the coupling constant and bare mass are fixed, see
Refs.~\cite{Schutz:1998jx,Krehl:1999km}. The $\Delta$ pole in the amplitude is generated by the $s$-channel
bare state, which couples to all the $I=3/2$ channels and the bare mass and coupling constants are fit parameters.
Meanwhile, the $u$-channel exchanged $\Delta$ is just regarded as a stable particle with the mass $1232$~MeV.
Because of numerical limitations, it is at the moment impossible to overcome these inconsistencies. 

\begin{figure}[t!]
	\centering
	\includegraphics[width=0.4\textwidth]{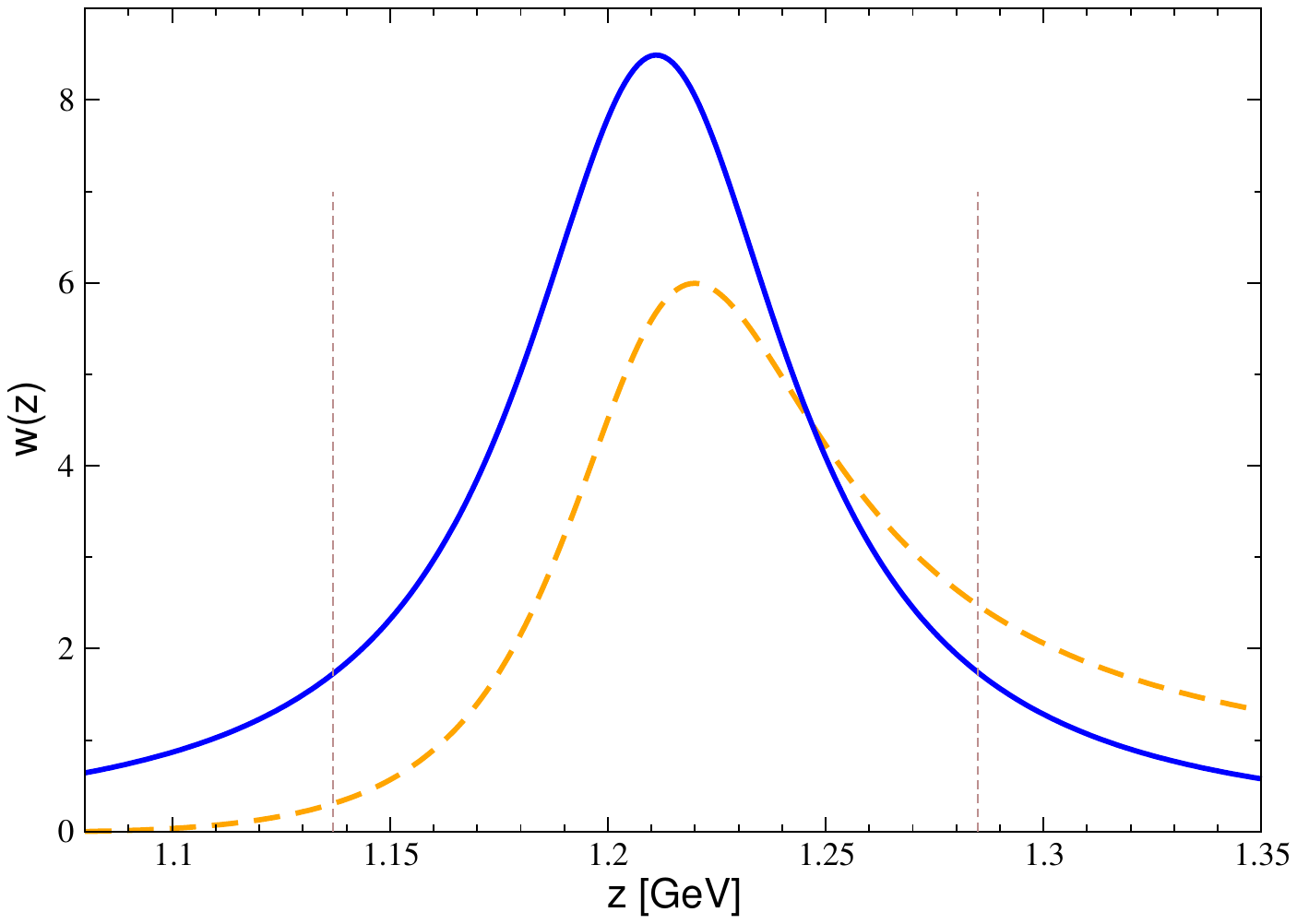}
	\caption{The spectral density functions the initial/final state $\Delta(1232)$. Blue solid line:
          the Breit-Wigner denominator in Eq.~\eqref{pedabww}. Orange dashed line: the spectral density
          function from the model. The vertical lines label the integral region of Eq.~\eqref{ZReval}. }
	\label{fig:Delif}
\end{figure}

Anyway, we may start the analyses with the initial/final state $\Delta$. Its bare mass in this model
is $M_0=1415$~MeV, with the coupling to the $\pi N$ channel $g^2/(4\pi)=0.36$. The resulting pole position
in the propagator of Eq.~\eqref{Gdef} is $1211-37i$~MeV. This is very similar to the single channel toy model,
and the spectral density function can be obtained directly from the imaginary part of Eq.~\eqref{Gdef},
see Fig.~\ref{fig:Delif}. The resulting elementariness is $69.6\%$, suggesting the elementary interpretation.
One can also calculate the complex elementariness of the Gamow state by directly applying Eq.~\eqref{pedaZgeneral}:
$Z=0.64-0.32i$ with the compositeness (only of $\pi N$ channel) $X=1-Z=0.36+0.32i$. Then the naive
measure of the elementariness is $59.5\%$, qualitatively in agreement with the spectral density function. 

As for the pole of $\Delta(1232)$ in the amplitude, there are two $s$-channel bare states in the $P_{33}$ wave,
and the spectral density functions show a typical mixture. The two structures, the first of which is the
$\Delta(1232)$, both show up in the first function. The elementariness directly from the model is $53.8\%$.
The local construction of Eq.~\eqref{Tcstsolve} fails. Similar to the $N^*(1675)$ state discussed in the last
section, here the $\pi\Delta$ residue is large and the imaginary part of the $\pi\Delta$ loop function
is positive. Then the whole imaginary part is positive, making the $g_1^2$ in Eq.~\eqref{Tcstsolve} negative.
If we use Eq.~\eqref{Nmres}, the constructed spectral density function gives an elementariness about $82.6\%$.
This ``plan B'' does not apparently violate the sum rule either: $\int_{m_\pi+m_N}^{\infty}w^{\text{lc}}(z)dz\simeq 0.93$.
Besides, this spectral density function from plan B does not show any negative value in the energy region we study. The spectral density function may suggest the $\Delta(1232)$ to be more elementary than composite, but on
the contrary the Gamow state has a large composition of the $\pi\Delta$ channel, leading to a rather small
measure of the elementariness, see Tab.~\ref{tab:selDX}. We cannot claim this is certainly true because of
the inconsistencies discussed above. 

\subsubsection*{\texorpdfstring{$\Delta(1600)\,\frac{3}{2}^+$}{}}
Though the $\Delta(1600)$ and the $\Delta(1232)$ are in the same wave, there are much less uncertainties
for the $\Delta(1600)$. In Fig.~\ref{fig:Dsdf}, it is seen that the peak of $\Delta(1600)$ mainly stays
on the first curve. The total elementariness from the two is rather moderate ($47.8\%$), while the local
construction succeeds and results in $77.5\%$. On the other side, as shown in Tab.~\ref{tab:selDX},
the biggest partial compositeness is of the $\pi\Delta$ channel, while the naive measure of the elementariness
is $69.7\%$. Combining the three results, the $\Delta(1600)$ is likely to be an elementary state. This is in accordance with the observation in Ref.~\cite{Ronchen:2022hqk} that the $\Delta(1600)$ is now largely induced by a bare $s$-channel pole, whereas it was dynamically generated in previous J\"uBo studies~\cite{Ronchen:2014cna,Ronchen:2015vfa,Ronchen:2018ury}.

\subsubsection*{\texorpdfstring{$\Delta(1700)\,\frac{3}{2}^-$}{}}
The result for the $\Delta(1700)$ in this study is interesting. The three scenarios agree with each other,
but all give rather moderate values. As seen from Fig.~\ref{fig:Dsdf}, the spectral density function from
the model and the local construction behave similarly. The model gives the value of the elementariness as
$59.7\%$, while the local construction results in is $44.9\%$. In Tab.~\ref{tab:selDX}, the naive measure
of the elementariness is also moderate ($47.8\%$), and the largest composition seems to be $\pi\Delta$.
One might claim that physically this state is really half-elementary and half-composite, but notice that
the width of the $\Delta(1700)$ resonance (almost $300$~MeV) is the biggest among all the selected states,
causing visible ambiguities. 

\section{Conclusion and outlook}\label{sec:con}
This paper studies the composition of the $N^*$ and $\Delta$ resonances, based on the coupled-channel
dynamics of the J{\"u}lich-Bonn model. Results from three scenarios are compared to draw the conclusions:
the spectral density functions directly from the model, the local construction from the pole parameters,
and the complex compositenesses. The first two scenarios stem from the physical scattering states,
which give probabilities as output, while the last is defined by the Gamow states and is different to the first
two in the basic philosophy. The three results roughly depict the model dependence of this study.
We have selected $13$ states, $8$ of which have reached relatively certain results: $N(1535) \frac{1}{2}^-$,
$N(1440) \frac{1}{2}^+$, $N(1710) \frac{1}{2}^+$, and $N(1520) \frac{3}{2}^-$ have chances of being composite,
whereas $N(1650) \frac{1}{2}^-$, $N(1900) \frac{3}{2}^+$, $N(1680) \frac{5}{2}^+$, and $\Delta(1600) \frac{3}{2}^+$
have tendencies to be elementary. For those $8$ states, at least two of the three scenarios result in
qualitative agreements. 

However, model uncertainties do exist. Some of the them are caused by the three-body effective channels,
especially the uncertainties of the $\Delta(1232)$ state is highly related to the $\pi\Delta$ channel. Including rigorous three-body unitarity and
crossing symmetry is a challenge left for future studies.  Moreover,
theoretically, the connections between the spectral density functions and the Gamow states still need to be
understood in a deeper way. Note that for $N(1650) \frac{1}{2}^-$, $N(1900) \frac{3}{2}^+$, and
$\Delta(1232) \frac{3}{2}^+$, the complex compositenesses of the Gamow states lead to opposite conclusions to
the spectral density functions. Moreover, relating the results which are based on a pure hadronic model to the descriptions in quark models appears to be difficult.
However, it is clear that a compact state in such an approach must have a mapping onto a quark model state, which we have assumed throughout. All in all, this topic itself is difficult in nature. One cannot expect
very clear interpretations and definite conclusions for every state, since only the states like the
deuteron can be highly related to physical observables. We believe this paper present the presently achievable status
of such type of investigation. We also hope this paper can trigger the community for further constructive
discussions on this topic.

In the future, the $\omega N$ channel can be included in the study of the compositions, after the $\omega N$
photoproduction is included in this model. Moreover, including more high-quality data also from other photoproduction channels will lead to refined pole values. The same methods can be applied on the other sectors,
e.g. the $\Lambda^*$ resonances or the $P_c$ exotic states, which are also being studied in the J{\"u}lich-Bonn model. 

\section*{Acknowledgements}
We thank Christoph Hanhart for many useful discussions and for participating in the early stages of this project. 
One of the authors (YFW) would like to thank Yu Lu for helpful discussions. The authors gratefully acknowledge
the computing time granted by the JARA Vergabegremium and provided on the JARA Partition part of the
supercomputer JURECA~\cite{JUWELS} at Forschungszentrum J{\"u}lich. This work is supported by the NSFC and
the Deutsche Forschungsgemeinschaft  (DFG, German Research Foundation) through the funds provided to the
Sino-German Collaborative Research Center TRR110 “Symmetries and the Emergence of Structure in
QCD” (NSFC Grant No. 12070131001, DFG Project-ID 196253076-TRR 110). Further support by the CAS
through a President’s International Fellowship Initiative (PIFI) (Grant No. 2018DM0034) and by the
VolkswagenStiftung (Grant No. 93562) is acknowledged.
This work is also supported by the MKW NRW under the funding code NW21-024-A and by the Deutsche Forschungsgemeinschaft (DFG, German Research Foundation) – 491111487.
\bibliographystyle{h-physrev}
\bibliography{mo_ref}

\end{document}